\documentclass[manuscript]{aastex}
\usepackage{multirow}
\usepackage{amsmath}
\usepackage{color}
\usepackage{ulem}

\slugcomment{}

\shorttitle{Optically thick HI dominant in the local interstellar medium}
\shortauthors{Y. Fukui et al.}

\begin{document}


\title{Optically thick H{\sc i} dominant in the local interstellar medium; an alternative interpretation to ``dark gas''\footnote{Based on observations obtained with {\it Planck} (http://www.esa.int/Planck), an ESA science mission with instruments and contributions directly funded by ESA Member States, NASA, and Canada.}
}


\author{Y. Fukui\altaffilmark{2}, K. Torii\altaffilmark{2}, T. Onishi\altaffilmark{3}, H. Yamamoto\altaffilmark{2}, R. Okamoto\altaffilmark{2}, T. Hayakawa\altaffilmark{2}, K. Tachihara\altaffilmark{2}, and H. Sano\altaffilmark{2}}
\affil{$^2$Department of Physics, Nagoya University, Chikusa-ku, Nagoya, Aichi 464-8601, Japan}
\affil{$^3$Department of Astrophysics, Graduate School of Science, Osaka Prefecture University, 1-1 Gakuen-cho, Nakaku, Sakai, Osaka 599-8531, Japan}
\email{fukui@a.phys.nagoya-u.ac.jp}

\begin{abstract}
Dark gas in the interstellar medium (ISM) is believed to not be detectable either in CO or H{\sc i} radio emission, but it is detected by other means including $\gamma$-rays, dust emission and extinction traced outside the Galactic plane at $|b|>5^\circ$. In these analyses, the 21-cm H{\sc i} emission is usually assumed to be completely optically thin. 
We have reanalyzed the H{\sc i} emission from the whole sky at $|b| > 15^\circ$ by considering temperature stratification in the ISM inferred from the {\it Planck}/{\it IRAS} analysis of the dust properties. 
The results indicate that the H{\sc i} emission is saturated with an optical depth ranging from 0.5 to 3 for 85\,\% of the local H{\sc i} gas. 
This optically thick H{\sc i} is characterized by spin temperature in the range 10\,K\,--\,60\,K, significantly lower than previously postulated in the literature, whereas such low temperature is consistent with emission/absorption measurements of the cool H{\sc i} toward radio continuum sources. 
The distribution and the column density of the H{\sc i} are consistent with those of the dark gas suggested by $\gamma$-rays, and it is possible that the dark gas in the Galaxy is dominated by optically thick cold H{\sc i} gas. 
This result implies that the average density of H{\sc i} is 2\,--\,2.5 times higher than that derived on the optically-thin assumption in the local ISM.

\end{abstract}

\keywords{ISM: clouds --- radiolines: ISM}

\section{Introduction}

It is important to quantify the constituents of the interstellar medium (ISM), which  mainly consists of neutral, molecular and ionized hydrogen H{\sc i}, H$_2$, and H{\sc ii}, in order to understand the role of the ISM in galactic evolution. The H{\sc i} gas has density mainly in a range from 0.01\,cm$^{-3}$ to 100\,cm$^{-3}$ while the CO probes the molecular hydrogen gas at density higher than 1000\,cm$^{-3}$, so the intermediate density regime 100\,cm$^{-3}$\,--\,1000\,cm$^{-3}$ may possibly remain unrecognized. It has been discussed that ``dark gas'' may exist, which is undetectable in radio emission, either in the 21-cm H{\sc i} or 2.6-mm CO transitions \citep{2005Sci...307.1292G,2011A&A...536A..19P}. Previous studies suggest that the dark gas probed by $\gamma$-rays and dust emission has a density regime between the H{\sc i} and H$_2$ as inferred from its spatial distribution intermediate between H{\sc i} and CO \citep{2005Sci...307.1292G,2011A&A...536A..19P}. 

The physical properties of the CO emitting molecular gas are relatively well understood due to transitions of its different rotational states and isotopic species, which allow us to derive physical and chemical parameters of the CO gas. 
On the other hand, the physical parameters of the H{\sc i} gas are more difficult to estimate, because the H{\sc i} line intensity is the only measurable quantity for a combination of two unknown parameters, the spin temperature $T_{\rm s}$ and optical depth $\tau_{\rm HI}$. 
The 21-cm transition is a transition characterized by the excitation temperature, called spin temperature, between the two spin-flip states in the electronic ground state. 
The H{\sc i} consists of warm neutral medium (WNM) and cold neutral medium \citep[CNM; for a review see][]{1990ARA&A..28..215D, 2009ARA&A..47...27K}. 
The mass of the H{\sc i} gas is measurable at reasonably high accuracy under the optically thin approximation, while the cold components having $T_{\rm s}$ of $\leq80$\,K may not be easily measurable because of optical depth effects. 
Only a comparison of the absorption and emission H{\sc i} profiles toward extragalactic radio continuum sources can be used to estimate $T_{\rm s}$ and $\tau_{\rm HI}$ \citep{2003ApJ...585..801D, 2003ApJ...586.1067H}, so the details of the cold H{\sc i} are still not fully understood. The existence of optically thick H{\sc i} in galaxies has been discussed based on line profiles, whereas a quantitative method to evaluate the physical properties has not yet been developed \citep{2012ApJ...749...87B}.

A recent work on the high latitude molecular clouds MBM\,53, 54, 55, and HLCG\,92-35 has shown that the H{\sc i} emission is optically thick in the surroundings of the CO clouds \citep[][Paper I]{2014ApJ...796...59F}. These authors compared the {\it Planck}/{\it IRAS} dust opacity \citep{2014A&A...571A..11P} with H{\sc i} and CO, and estimated $T_{\rm s}$ to be 20\,K\,--\,40\,K (average is 30\,K) and $\tau_{\rm HI}$ 0.3\,--\,5 (average is 2) by assuming that the dust opacity is proportional to the ISM proton column density. They suggest that the H{\sc i} envelope is massive having more than 10 times the mass of the CO clouds, and that such optically thick H{\sc i} may explain the origin of the dark gas, an alternative to CO-free H$_{2}$. It is important to test if the H{\sc i} shows similar high optical depth in a much larger portion of the sky. 

In addition, it is notable that in three TeV $\gamma$-ray supernova remnants RX\,J1713.7-3946, RX\,J0852.0-4622, and HESS\,J1731-347, it is found that spatially extended cold H{\sc i} gas of $T_{\rm s} \sim 40$\,K which has no CO emission is responsible for the $\gamma$-rays via the hadronic process between the cosmic-ray protons and the interstellar protons \citep{2012ApJ...746...82F, 2012IAUS..284..389T,2013ASSP...34..249F,Fukuda2014}. The cold H{\sc i} probably represents the compressed H{\sc i} shell swept up by the stellar winds of the supernova progenitor. This finding raised independently a possibility that the cold H{\sc i} gas may be more ubiquitous than previously thought. 

It has been difficult to derive $T_{\rm s}$ and $\tau_{\rm HI}$ in general and our knowledge on the cold H{\sc i} remains ambiguous at best. In order to better understand the relationship between the dust emission and H{\sc i} over a significant portion of the sky, we have compared the H{\sc i} with the dust properties derived from the {\it Planck} and {\it IRAS} data beyond the area studied by Paper I. This comparison was made by using the H{\sc i} dataset at $33'$ resolution from the Leiden/Argentine/Bonn (LAB) archive data and the {\it Planck}/{\it IRAS} dust properties. We present the results of the detailed comparison. Section 2 presents the observations, Section 3 the results, Section 4 the discussion and Section 5 our conclusions.

\section{Observational Datasets}
\subsection{Datasets}
In this study we used the all sky maps of the dust model data measured by {\it Planck}/{\it IRAS}, the LAB H{\sc i} data, the CfA CO data, 1.4\,GHz radio continuum data, and H$\alpha$ data. 
All the datasets are smoothed to be a HPBW of 33$'$, which corresponds to the HPBW of the LAB H{\sc i} data, and are then converted into the Mollweide projection with a grid spacing of 30$'$.

\subsubsection{H{\sc i} Data}
The LAB H{\sc i} 21-cm survey \citep{2005A&A...440..775K} is used in this study. It covers the entire sky at an effective angular resolution of 33$'$ (HPBW). The data is taken from the web page of the LAMBDA project\footnote[4]{http://lambda.gsfc.nasa.gov} with Healpix format\footnote[5]{http://healpix.jpl.nasa.gov}. 
The rms noise fluctuations of the H{\sc i} data are 0.07\,K\,--\,0.09\,K in $T_{\rm mb}$ at 1\,km\,s$^{-1}$ velocity resolution \citep{2005A&A...440..775K}.
%All the other datasets are spatially smoothed to the same spatial resolution and then are converted into the same projection. 
The effective velocity range of the present analysis is $\sim\pm10$\,km\,s$^{-1}$ in $v_{\rm LSR}$ where the H{\sc i} is peaked at around 0\,km\,s$^{-1}$, while the H{\sc i} integrated intensity $W_{\rm HI}$ is calculated using a radial velocity, $v_{\rm LSR}$, range from $-150$\,km\,s$^{-1}$ to $+150$\,km\,s$^{-1}$.  
More details on the velocity range are given in Section 2.2.

\subsubsection{Planck/IRAS Data}
Archival datasets of dust optical depth at 353\,GHz, $\tau_{353}$, and dust temperature, $T_{\rm d}$, are used to make comparisons with the H{\sc i} data, where the 353, 545 and 857\,GHz data of the first 15 months observations with {\it Planck} and the 100\,$\mu$m data obtained with {\it IRAS} are used to derive $\tau_{353}$ and $T_{\rm d}$ \citep[see][for details]{2014A&A...571A..11P}.
Here we utilize version R1.20 of the {\it Planck} maps.

\subsubsection{CO Data}
We use the integrated intensity $^{12}$CO $J$=1--0 map over the full observed velocity range by \citet{2001ApJ...547..792D} as a molecular gas tracer. The HPBW is 8.7$'$, and the rms noise fluctuations are 0.1\,K\,--\,0.35\,K at 1.3\,km\,s$^{-1}$ velocity resolution \citep{2001ApJ...547..792D}. The data are also taken from the LAMBDA project page.

\subsubsection{Background 21-cm Continuum Data}
The 21-cm continuum emission is used as background emission of H{\sc i}. We use the CHIPASS 21-cm radio continuum map with a sensitivity of $\sim$40\,mK \citep{2014PASA...31....7C} for the southern sky and the Stockert 21-cm radio continuum map with a sensitivity of $\sim$50\,mK \citep{1986A&AS...63..205R} for the northern sky. The original HPBWs are $14.4'$ and $35'$, respectively. This continuum emission includes the cosmic microwave background emission.

\subsubsection{H$\alpha$ Data}
In order to mask the region where dust is locally heated or destroyed by ultraviolet (UV) radiation, we use the all-sky H$\alpha$ data provided by \citet{2003ApJS..146..407F}. The angular resolution is 6$'$. Typical uncertainties in each pixel is estimated to be 0.3\,--1.3\,Rayleigh for $|b|>15^\circ$ and 1\,--\,5\,Rayleigh for $|b| \leq 15^\circ$. 

\subsection{Masking}

\begin{figure}
\epsscale{.80}
\plotone{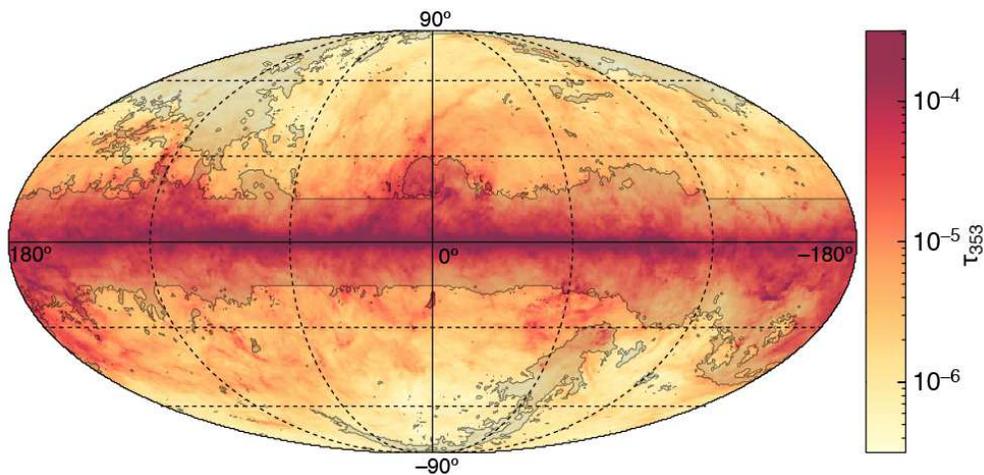}
\caption{
All-sky Mollweide projections of $\tau_353$ distribution in the galactic coordinate. The center of the map is $(l, b)=(0^\circ, 0^\circ)$, and the coverages of l and b are from $-180^\circ$ to $+180^\circ$ and from $-90^\circ$ to $+90^\circ$, respectively. Dashed lines are plotted every $60^\circ$ in $l$ and every $30^\circ$ in $b$. The masked region used in the present analysis is shown by shading.
 \label{tau353_allsky}} 
\end{figure}

Figure \ref{tau353_allsky} shows the distribution of $\tau_{353}$, which includes mainly local clouds like Taurus, Lupus, Aquila, Polaris flare and Chameleon within 200\,pc of the Sun. 
A correlation analysis between dust and gas must be done toward a single component of the ISM and regions that do not overlap along the line of sight are chosen for comparison. Otherwise, regions of different physical properties will be mixed up, reducing the correlation among the physical parameters. It is also important to avoid contamination so that local irradiation by high-mass stars does not significantly affect the dust emission. Considering these, we selected the region for the present analysis according to the following five criteria.

\begin{figure}
\epsscale{.90}
\plotone{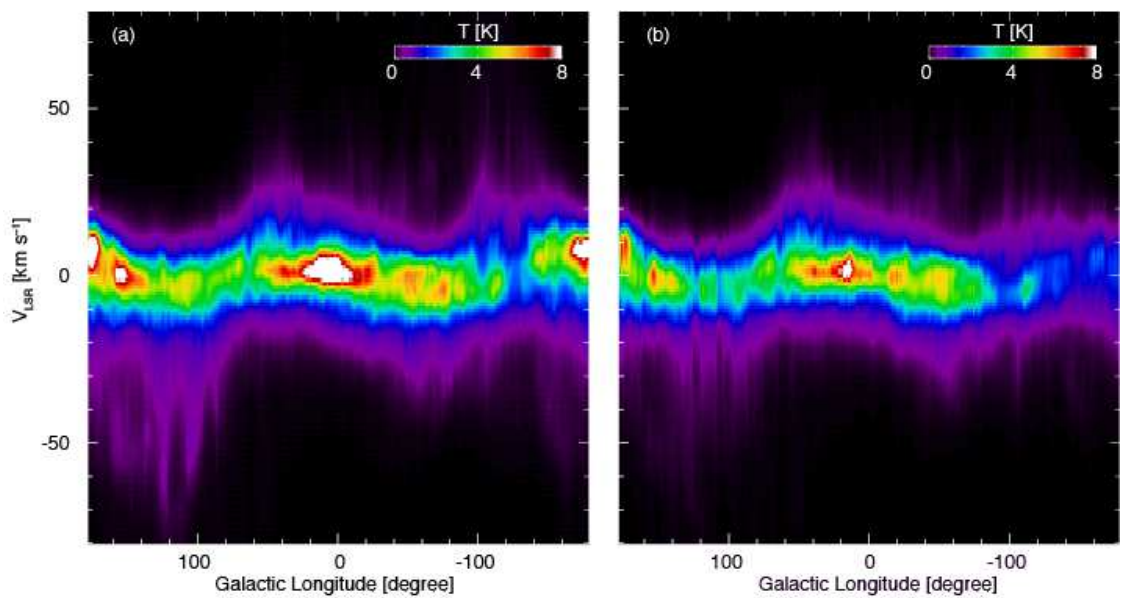}
\caption{
Longitude-velocity distributions of latitude averaged all-sky H{\sc i} intensity after masking are shown. (See text in Section 2.2). (a) Criteria 1, 2, 4 and 5 are applied. (b) All the criteria are applied. \label{HI_lv}} 
\end{figure}

\begin{enumerate}

\item The Galactic latitude $|b|$ is higher than 15$^\circ$ in order to eliminate contamination by different components in the Galactic plane along the line of sight.
At intermediate latitudes the assumption that the ISM emission in any given direction is dominated by a single structure (cloud or cloud complex) is a valid approximation over most of the sky.  
This approximation ignores the very diffuse gas (inter-cloud medium) that is present at some level in all directions.  At high latitudes the diffuse gas (neutral and ionized) is sometimes the only form of interstellar matter on the line of sight, with column density below about 10$^{20}$\,cm$^{-2}$.  There is no evidence for ``dark gas'' at these low column densities, so we ignore the diffuse gas in this analysis.
\item Extragalactic objects such as the LMC, SMC and M\,31 are removed from the present analysis. The points where $W_{\rm HI}$ at $|v_{\rm LSR}|>100$\,km\,s$^{-1}$ are larger than 10 \% of those at $|v_{\rm LSR}|<70$\,km\,s$^{-1}$ are masked.
In addition, the low velocity component of the Magellanic stream seen at $l \sim -60^\circ$\,--\,$+60^\circ$ and $b < -60^\circ$ is masked by hand using the integrated intensity map at $|v_{\rm LSR}|<70$\,km\,s$^{-1}$ to show a continuous distribution with the components seen at $|v_{\rm LSR}|>70$\,km\,s$^{-1}$.
\item As seen in the H{\sc i} longitude-velocity map in Figure \ref{HI_lv}(a), the H{\sc i} components are concentrated around $|v_{\rm LSR}|<30$\,km\,s$^{-1}$. We thus remove the points that have H{\sc i} emission from $-70$\,km\,s$^{-1}$ to $-35$\,km\,s$^{-1}$, those are identified as the intermediate velocity clouds mainly distributed at $60^\circ < b < 80^\circ$ \citep{2011A&A...536A..24P,2014A&A...571A..11P}. The points where $W_{\rm HI}$ are larger than 50\,K\,km\,s$^{-1}$ at 35\,km\,s$^{-1}$$ < |v_{\rm LSR}| < $70\,km\,s$^{-1}$ are masked.
\item In order to eliminate locally heated components, the H$\alpha$ emission is restricted to be weaker than 5 Rayleigh \citep{2003ApJS..146..407F}. This eliminates the dust in ionized regions (e.g., in Orion A, the Ophiuchus region, etc.), where the dust may be destroyed or heated by the UV photons, causing an anomalous dust to gas ratio and dust temperature.
\item CO gas has a different physical environment with higher density. Regions with detectable CO emission with integrated intensities of larger than 1\,K\,km\,s$^{-1}$ are excluded by using the CO data provided by \citet{2001ApJ...547..792D}. Individual regions of CO gas not included in the present study will be dealt with in detail at higher resolution in separate papers. 

\end{enumerate}

\begin{figure}
\epsscale{.90}
\plotone{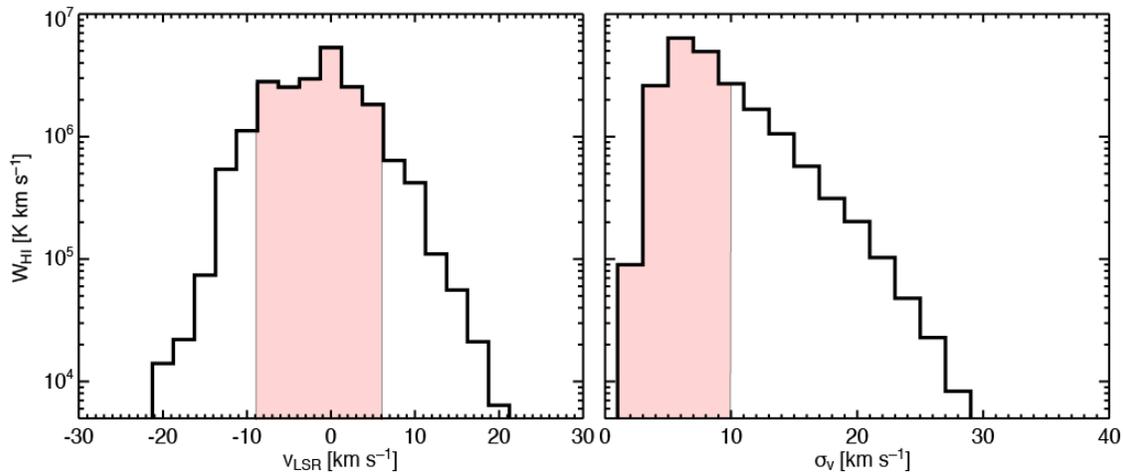}
\caption{
Intensity weighted histograms of (a) the peak velocity and (b) 1\,$\sigma$ velocity dispersion, $\sigma_v$, of the H{\sc i} emission shown in Figure \ref{HI_lv}(b).
The shaded areas in (a) and (b) show the velocity ranges that account for 86\,\% and 81\,\% of $W_{\rm HI}$, respectively.
 \label{hist_v}} 
\end{figure}

The entire masked region is shown in Figure \ref{tau353_allsky}, and Figure \ref{HI_lv}(a) shows the longitude-velocity map after masking by criteria 1, 2, 4 and 5, and Figure \ref{HI_lv}(b) by all the criteria including the criterion 3. 
Figure \ref{hist_v} also shows the histograms of the peak velocities and 1\,$\sigma$ velocity dispersions, $\sigma_v$, for the resulting H{\sc i} data.
The present analysis substantially excludes, by masking, components that are broad and separated by more than 35\,km\,s$^{-1}$ from the main H{\sc i} emission at around 0\,km\,s$^{-1}$ (Figure \ref{hist_v}). 
The fraction of $W_{\rm HI}$ of $v_{\rm LSR}$ between $-9$\,km\,s$^{-1}$ and $+6$\,km\,s$^{-1}$ accounts for 86\,\% (Figure \ref{hist_v}(a)) and that having $\sigma_v$ less than 10\,km\,s$^{-1}$ is 81\,\%  (Figure \ref{hist_v}(b)), while we otherwise estimate $W_{\rm HI}$ by using a $v_{\rm LSR}$ range from $-150$\,km\,s$^{-1}$ to $+150$\,km\,s$^{-1}$.
The main H{\sc i} component corresponds to velocities showing CNM in general. 
On the other hand, the major part of the WNM has large linewidth of more than 20\,km\,s$^{-1}$ and a peak velocity often shifted by more than 10\,km\,s$^{-1}$ from the main H{\sc i} peak \citep{2003ApJS..145..329H}. 
A single $T_{\rm s}$ for the present substantially narrow velocity range is a reasonable assumption as a first-order approximation, and is consistent with the results in the high-latitude clouds (Paper I).

\section{Results}

\subsection{$\tau_{353}$ Versus $N_{\rm HI}$}

\begin{figure}
\epsscale{.50}
\plotone{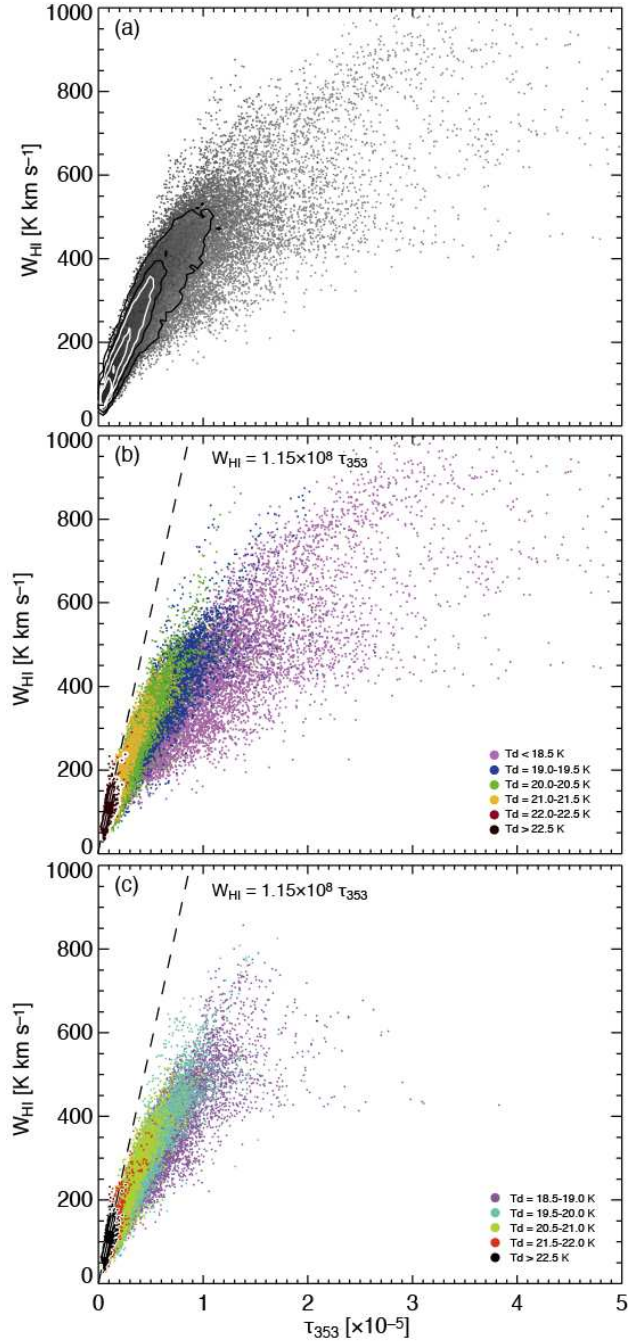}
\caption{
(a) Scatter plot between $\tau_{353}$ and $W_{\rm HI}$. Contours are plotted at 1\,\%, 5\,\%, 10\,\%, 30\,\%, and 50\,\% of the peak.
(b, c) The scatter plots of $\tau_{353}$ and $W_{\rm HI}$ for $T_{\rm d}$ in windows of 0.5\,K interval every 1\,K.
 \label{tau353_vs_whi}} 
\end{figure}

Figure \ref{tau353_vs_whi}(a) shows a scatter plot between $W_{\rm HI}$ and $\tau_{353}$ for the entire region analyzed. This plot has a correlation coefficient of 0.79. In the same manner in Paper I, in order to see the dependence on $T_{\rm d}$, we colored the data points in a window of 0.5\,K in $T_{\rm d}$ every 1\,K in Figures \ref{tau353_vs_whi}(b) and \ref{tau353_vs_whi}(c). We see clearly that points for each $T_{\rm d}$ show better correlation with a correlation coefficient higher than 0.9. 
We obtained best-fit straight regression lines by least-squares fitting (Table \ref{table_slope}). 
Here, for the ranges 22.0\,K\,$\leq \ T_{\rm d}\,<\,$\,22.5\,K and 22.5\,K\,$\leq\,T_{\rm d}$, the number of points is less than for the other ranges, and we thus assume that the intercept is zero and make fits only for slopes.
The regression shows a trend that the slope of $W_{\rm HI}$ with respect to $\tau_{353}$ becomes smaller systematically with decreasing $T_{\rm d}$. This trend is similar to what is found in the case of high-latitude clouds (Paper I). 
Recent studies by the {\it Planck} Collaboration have also found that $T_{\rm d}$ increases with decreasing gas column density \citep{2014A&A...571A..11P, 2014A&A...566A..55P}.
By following the same argument as in Paper I, we expect that the H{\sc i} emission is moderately optically thick in general and that the saturation due to the optical depth effect causes weaker $W_{\rm HI}$ for lower $T_{\rm d}$ and higher $\tau_{353}$ . The optically thin limit is seen only for the highest $T_{\rm d}$, and the rest of the points with lower $T_{\rm d}$ and shallower slopes suffer from saturation in H{\sc i} intensity. We apply the optically thin relationship in equation (\ref{eq1}) in order to convert $W_{\rm HI}$ into the H{\sc i} column density under the optically thin approximation $N_{\rm HI}^*$,

\begin{equation}
N_{\mathrm{HI}}^*\,[\mathrm{cm^{-2}}]\,=\,1.823\times10^{18}\cdot W_{\mathrm{HI}}\,[\mathrm{K\,km\,s^{-1}}],\label{eq1}
\end{equation}

With the optically thin points at 22.5\,K $\leq \ T_{\rm d}$ in Figures \ref{tau353_vs_whi}(b) and (c), we obtain a relationship between $W_{\rm HI}$ and $\tau_{353}$ 

\begin{equation}
W_{\mathrm{HI}}\,[\mathrm{K\,km\,s^{-1}}]= k\cdot \tau_{353} \label{eq2}, 
\end{equation}
where $k=1.15\times10^8$ K\,km\,s$^{-1}$ is a constant. 
This relation will hold for any optically thin values of $\tau_{353}$, even if the H{\sc i} emission is not optically thin, as long as the dust properties are uniform with no significant spatial variation. We tested the variation of $k$ as a function of $b$ and find that the peak-to-peak dispersion of $k$ is less than 10\,\% for $|b|=50^\circ$\,--\,90$^\circ$.
Finally using equations (\ref{eq1}) and (\ref{eq2}), we can convert $\tau_{353}$ into the H{\sc i} column density, $N_{\rm HI}$, by the following equation,
\begin{equation}
\frac{N_{\rm HI}\,[\mathrm{cm^{-2}}]}{N_0} = \frac{\tau_{353}}{\tau_0}, \label{eq2+}
\end{equation}
where $N_0 = 1\times10^{21}$ cm$^{-2}$ and $\tau_0 = 4.77 \times 10^{-6}$ are typical values for intermediate latitude lines of sight (see Section 3.2 and Figure \ref{hist_ts} below).

\subsection{Temperature-dependent Analysis of H{\sc i}}

The above results suggest that the H{\sc i} emission is partially saturated due to the optical depth effect for lower temperatures and the degree of saturation depends on $T_{\rm s}$ of the H{\sc i} in a way similar to the high-latitude clouds (Paper I). $W_{\rm HI}$ is expressed by a radiation transfer equation

\begin{equation}
W_{\mathrm{HI}}\,[\mathrm{K\,km\,s^{-1}}]= (T_{\mathrm{s}}\,[\mathrm{K}] - T_{\mathrm{bg}}\,[\mathrm{K}])\cdot \Delta V_{\mathrm{HI}}\,[\mathrm{km\,s^{-1}}] \cdot [1-\exp(-\tau_{\mathrm{HI}})] \label{eq3}, 
\end{equation}
where $T_{\rm s}$ is assumed to be uniform on the line of sight, $T_{\rm bg}$ is the background radio continuum radiation, and $\Delta V_{\rm HI}$ is the H{\sc i} linewidth ($\Delta V_{\rm HI}=W_{\rm HI}/T_{\rm peak}$ for single-component spectra, where $T_{\rm peak}$ is the peak temperature of the H{\sc i} emission at each point). 
Equation 4 makes the assumption that all the H{\sc i} in each velocity channel is at a single temperature, and $\tau_{\rm HI}$ is the average in $\Delta V_{\rm HI}$.

By assuming that $\tau_{353}$ is solely ascribed to the dust in the H{\sc i} gas, we are able to estimate $T_{\rm s}$ and  $\tau_{\rm HI}$. This assumption needs to be examined because molecular hydrogen with no detectable CO may account for some fraction of the total column density, $N_{\rm H}$, at space density, $n_{\rm H}$ above 100\,cm$^{-3}$. It is also worth considering whether the spatial variation of dust properties may offer an alternative explanation. We shall discuss these possibilities later in Section 4. 

\begin{figure}
\epsscale{.5}
\plotone{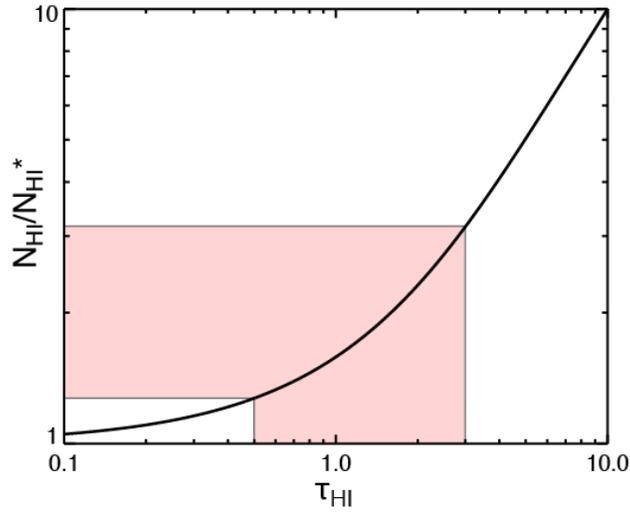}
\caption{
Curve of equation (\ref{eq4}), $N_{\mathrm{HI}}/N_{\mathrm{HI}}^*= \tau_{\mathrm{HI}}/[1-\exp(-\tau_{\mathrm{HI}})]$, is shown. The shaded area shows the typical $\tau_{\rm HI}$ range of 0.5\,--\,3.0 determined in Figure \ref{hist_ts}(a), which correspond to $N_{\rm HI}$/$N_{\rm HI}^*$ of $\sim$1.3\,--\,3.1, respectively.
 \label{figure_eq4}} 
\end{figure}

Under the present assumption, $\tau_{353}$ is converted into $N_{\rm HI}$ by equation (\ref{eq2+}) or $N_{\rm HI}\,{\rm [cm^{-2}]} = 2.1\times10^{26} \cdot \tau_{353}$, and $W_{\rm HI}$ into $N_{\rm HI}$* by equation (\ref{eq2}). $N_{\rm HI}^*$ gives an underestimate of the actual $N_{\rm HI}$ if the H{\sc i} emission is not optically thin, and the ratio $N_{\rm HI}$/$N_{\rm HI}^*$ is given as 
\begin{equation}
N_{\mathrm{HI}}/N_{\mathrm{HI}}^*= \tau_{\mathrm{HI}}/[1-\exp(-\tau_{\mathrm{HI}})] \label{eq4},
\end{equation}
where $\tau_{\rm HI}$ is given by the following equation which is valid for any positive optical depth; 
\begin{equation}
\tau_{\mathrm{HI}}=\frac{N_{\mathrm{HI}}\,[\mathrm{K\,km\,s^{-1}}]}{1.823\times10^{18}} \cdot \frac{1}{T_{\mathrm s}\,[\mathrm{K}]} \cdot  \frac{1}{\Delta V_{\mathrm{HI}}\,[\mathrm{km\,s^{-1}}]} \label{eq5}.
\end{equation}
Figure \ref{figure_eq4} shows a curve of equation (\ref{eq4}).

Along with the assumption that there is only one cloud structure that dominates the emission by dust and gas on any given line of sight,
we further assume that there is a single $T_{\rm s}$ for the H{\sc i} gas in this structure.  
Relaxing this assumption to include warm and cool gas phases associated with the cloud would make equations (\ref{eq3}) and (\ref{eq5}) more complicated, and introduce more unknown quantities into the analysis. 
For the present paper, we keep the analysis simple by assuming that a single spin temperature dominates the H{\sc i} on each line of sight.

\begin{figure}
\epsscale{1.}
\plotone{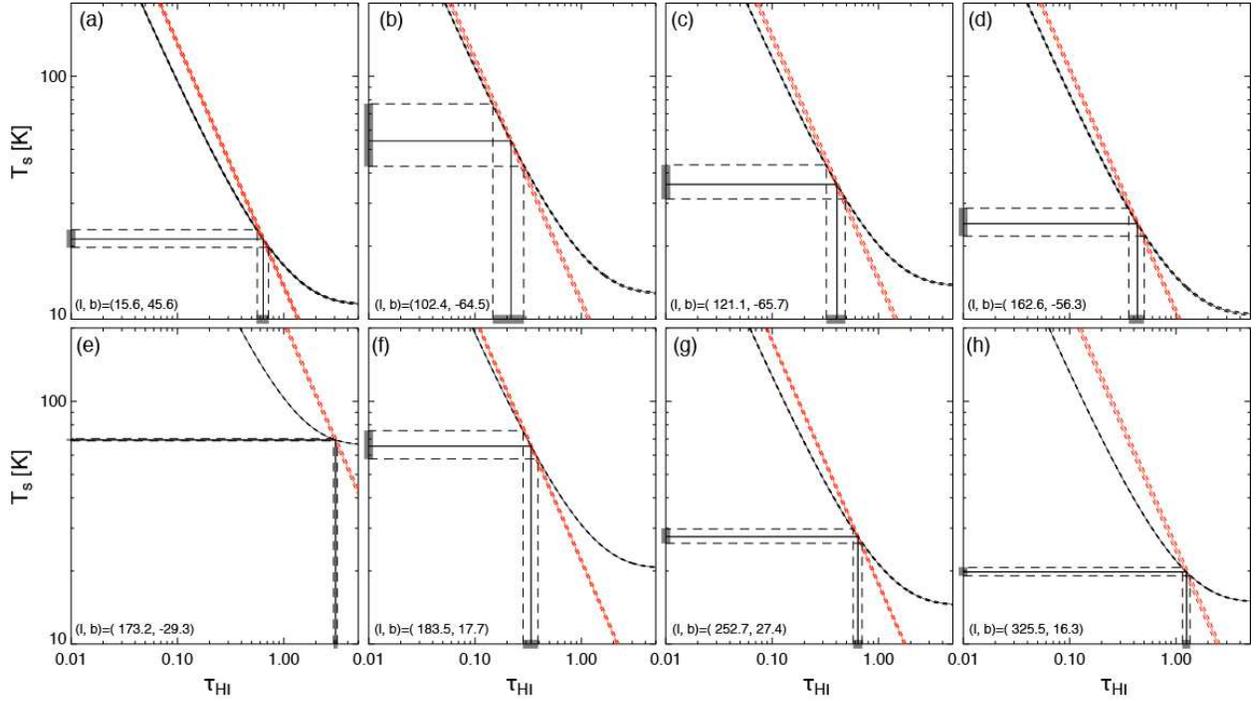}
\caption{
Examples for $\tau_{\rm HI}$ and $T_{\rm s}$ estimates. Black line shows curves given by equation (\ref{eq3}), while red line equation (\ref{eq5}). Dotted lines show the error range of each curve, where the intrinsic observational errors in $\tau_{353}$ and $W_{\rm HI}$ are considered. The resulting errors for $\tau_{\rm HI}$ and $T_{\rm s}$ are shown with thick line on the bottom and left axes of each panel.
 \label{figure_fitting}} 
\end{figure}

\begin{figure}
\epsscale{1.}
\plotone{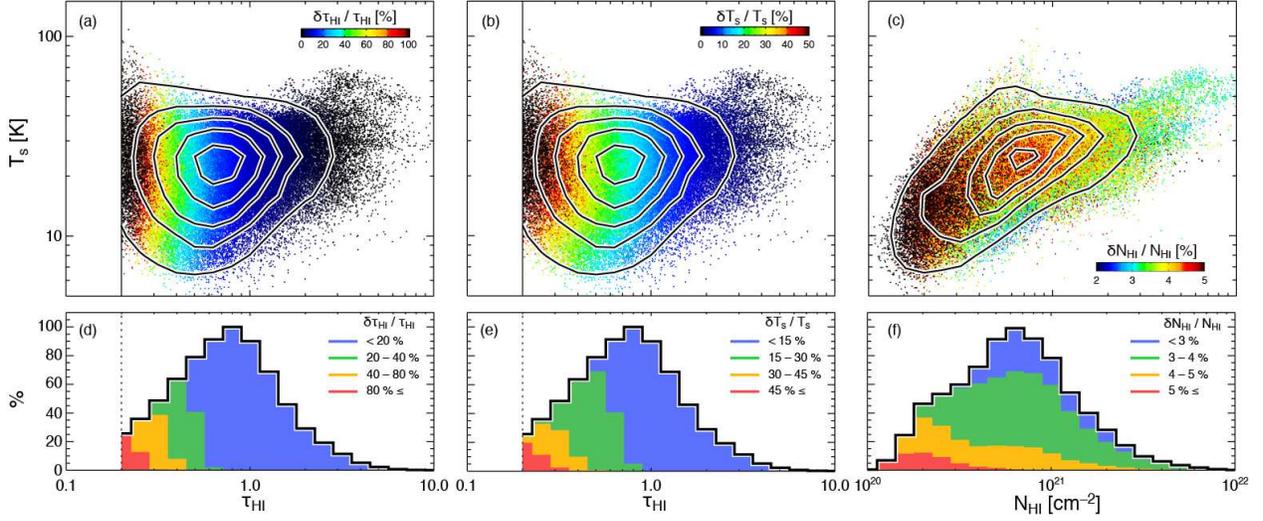}
\caption{
(a, b) Distributions of $\tau_{\rm HI}$ and $T_{\rm s}$, where color represents the fractional errors of (a) $\tau_{\rm HI}$ and (b) $T_{\rm s}$ (in percent). The errors indicate the full lengths of the bars at 1\,$\sigma$ levels in Figure \ref{figure_fitting}.
(c) Distributions of $N_{\rm HI}$ and $T_{\rm s}$. Color represents the fractional 1\,$\sigma$ error of $N_{\rm HI}$.
Contours show the 2D histograms and are plotted at 10\,\%, 30\,\%, 50\,\%, 70\,\%, and 90\,\% of the peaks. 
(d--f) Stacked histograms of $\tau_{\rm HI}$ for various $\tau_{\rm HI}$ and $T_{\rm s}$ error ranges, respectively. 
(f) Stacked histograms of $N_{\rm HI}$ for various $N_{\rm HI}$ error ranges. 
Each histogram in (d--e) is normalized at its peak.
 \label{err_plot}} 
\end{figure}

We here solve the two coupled equations (\ref{eq3}) and (\ref{eq5}) to estimate $T_{\rm s}$ and $\tau_{\rm HI}$. Note that these two equations are independent as long as $\tau_{\rm HI}$ is finite, although they become essentially a single equation in the optically thin limit. 
Discussion of optically thick H{\sc i} is also found in the literature \citep{2004ApJ...603..560S,2013pss5.book..549D}.
Examples for $T_{\rm s}$ and $\tau_{\rm HI}$ determination are shown in Figure \ref{figure_fitting}, with the results summarized in Table \ref{table_fitting}. 
Figure \ref{err_plot} shows the distributions of $T_{\rm s}$, $\tau_{\rm HI}$ and $N_{\rm HI}$ colored by their fractional errors. 
Here we estimate the error in $T_{\rm s}$ and $\tau_{\rm HI}$ as shown by thick lines in Figure \ref{figure_fitting} from the 1\,$\sigma$ uncertainties of $\tau_{353}$ and $W_{\rm HI}$.
The error in $N_{\rm HI}$ is also calculated using equation (\ref{eq2+}) from the error in $\tau_{353}$.
It is shown that the method cannot be applied with high accuracy when $\tau_{\rm HI}$ is smaller than $\sim$0.2 because of the degeneracy of equations (\ref{eq3}) and (\ref{eq5}). We therefore excluded the regions where $\tau_{\rm HI}$ is smaller than 0.2. 
Figure \ref{err_plot}(c) indicates that $N_{\rm HI}$ is accurately determined within $\pm5$\,\%, while the errors in $T_{\rm s}$ and $\tau_{\rm HI}$ are not always small.

\begin{figure}
\epsscale{.7}
\plotone{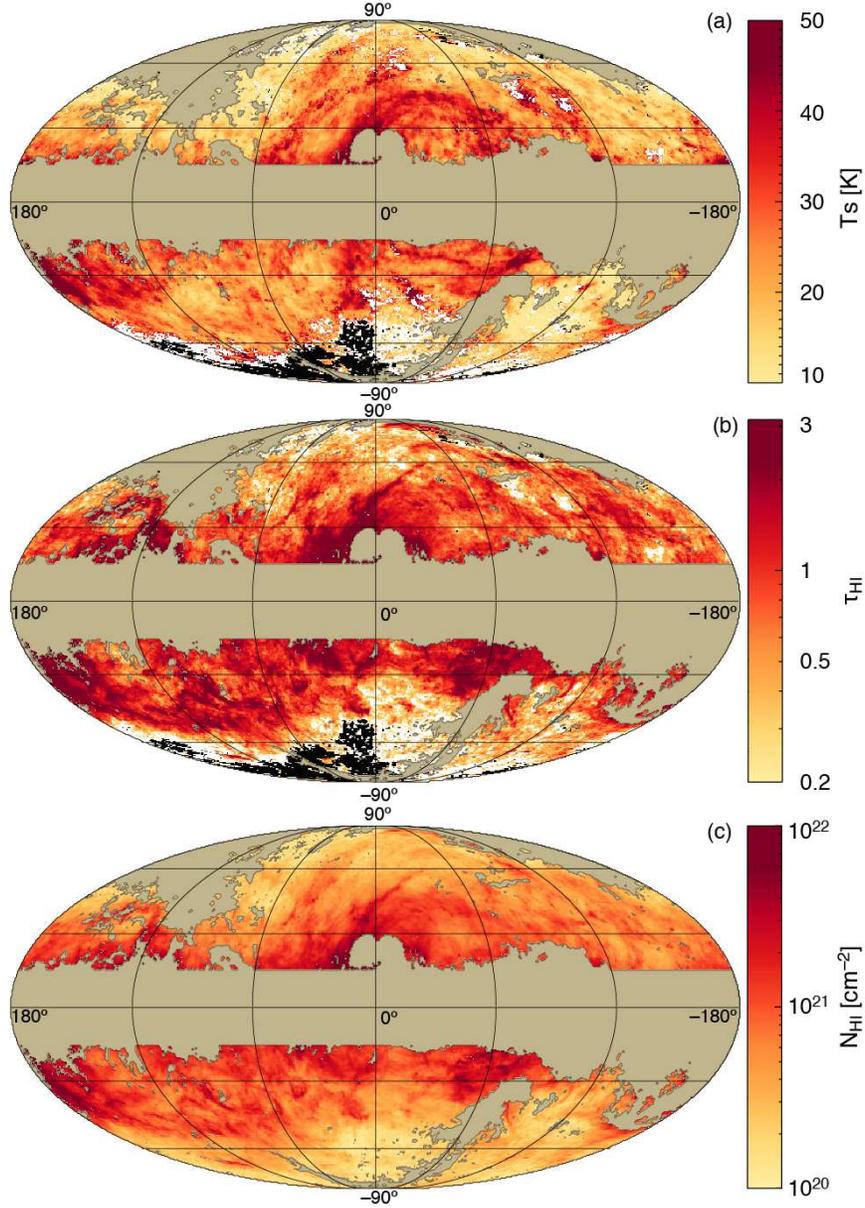}
\caption{
All-sky distributions of the (a) $T_{\rm s}$ and (b) $\tau_{\rm HI}$ maps. 
The masked region is shown in gray.
The black dots represent the region where $T_{\rm d} > 22.5$ K. 
The white dots show the points where $T_{\rm s}$ and $\tau_{\rm HI}$ are not determined because $\tau_{\rm HI} < 0.2$.
In (c) we use $N_{\rm HI}^*$ instead of $N_{\rm HI}$ for the points colored in white or black in (a) and (b).
The center of the map is $(l, b)=(0^\circ, 0^\circ)$, and the coverages of l and b are from $-180^\circ$ to $+180^\circ$ and from $-90^\circ$ to $+90^\circ$, respectively. Dashed lines are plotted every $60^\circ$ in $l$ and every $30^\circ$ in $b$. 
 \label{ts_allsky}} 
\end{figure}

\begin{figure}
\epsscale{1.}
\plotone{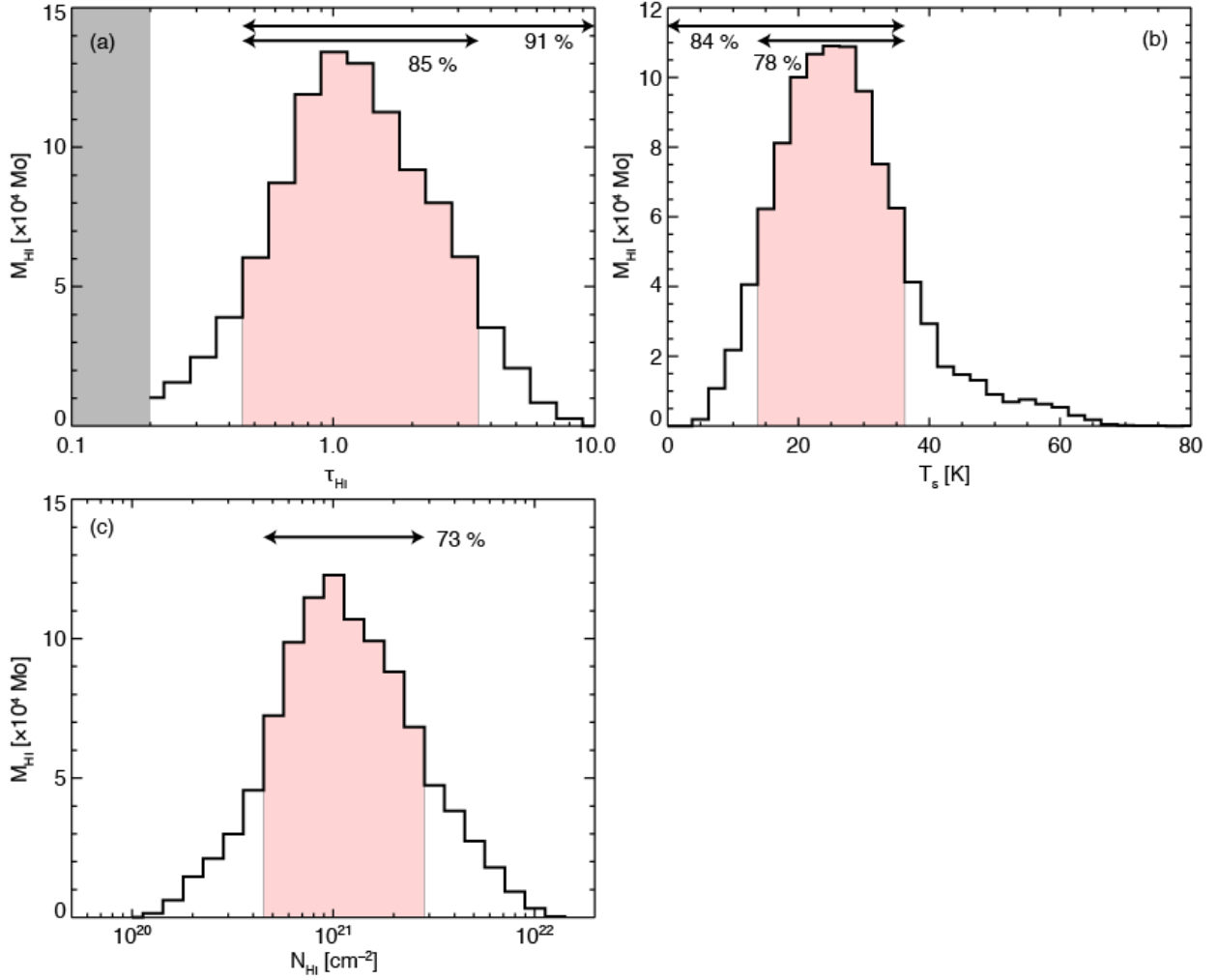}
\caption{
Histograms of (a) $T_{\rm s}$, (b) $\tau_{\rm HI}$ and (c) $T_{\rm d}$  weighted by H{\sc i} mass, where distance is assumed to be 150\,pc. The gray area in (a) indicates the lower limit of $\tau_{\rm HI}$.
The area filled in pink in each panel is defined to contain about 70\,\%\,--\,80\,\% of the all points.
 \label{hist_ts}} 
\end{figure}

The spatial distributions of the derived $T_{\rm s}$, $\tau_{\rm HI}$ and $N_{\rm HI}$ are shown in Figure \ref{ts_allsky}, and Figure \ref{hist_ts} shows histograms of these three parameters.
In Figures\,\ref{ts_allsky}(a) and (b) the points with $\tau_{\rm HI}$ $<$ 0.2 are shown in white, and we use $N_{\rm HI}^*$ instead of $N_{\rm HI}$ for these points in Figure\,\ref{ts_allsky}(c), while these points are not included in Figure\,\ref{hist_ts}. 
The mass ratio of the points with $\tau_{\rm HI} < 0.2$ to all the data points is only 3\,\%.
As seen in Figure \ref{hist_ts}, $\tau_{\rm HI}$ ranges from 0.5 to 3.0 for 85\,\% of the total H{\sc i} gas, where the H{\sc i} gas with $\tau_{\rm HI}$ larger than 0.5 accounts for 91\,\%.
On the other hand, $T_{\rm s}$ ranges from 15\,K to 35\,K for 78\,\% and $T_{\rm s}$ is less than 35\,K for 84\,\%.
$N_{\rm HI}$ ranges from $5\times10^{20}$\,cm$^{-2}$ to $3\times10^{21}$\,cm$^{-2}$ for 73\,\%, where the peak is seen at $10^{21}$\,cm$^{-2}$.

There has been discussion that $T_{\rm s}$ is generally higher than 80\,K, with 130\,K as the nominal value in the literature \citep[e.g.,][]{2004JApA...25..185M}. On the other hand, \citet{2003ApJ...585..801D} and \citet{2003ApJ...586.1067H} showed that there exists cold H{\sc i} gas (CNM) having $T_{\rm s}$ of 20\,K\,--\,50\,K from H{\sc i} emission/absorption profiles toward radio continuum background sources. This $T_{\rm s}$ range is consistent with the current $T_{\rm s}$ distribution. 

\begin{figure}
\epsscale{.7}
\plotone{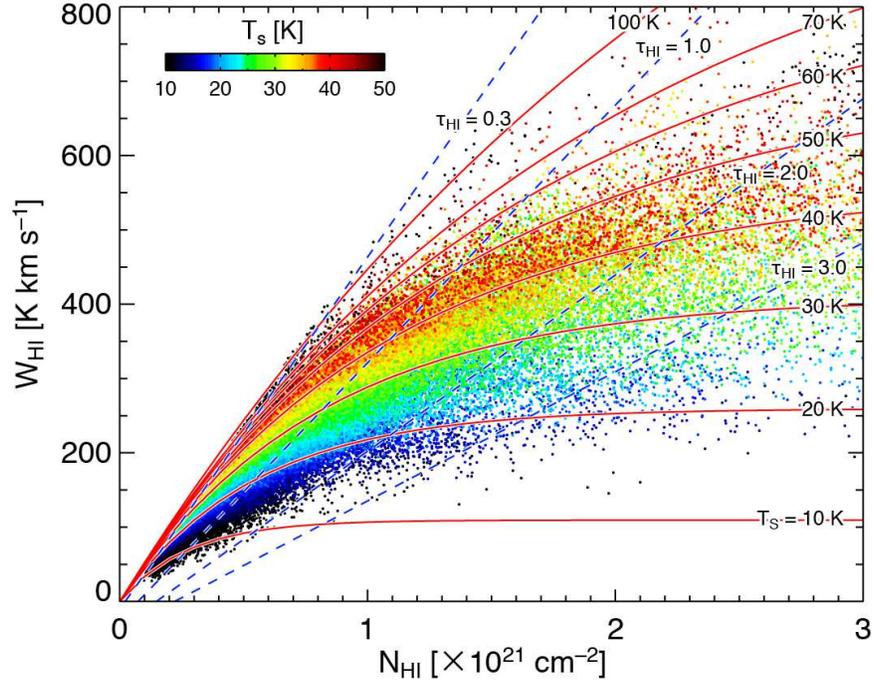}
\caption{
Correlation plot between $W_{\rm HI}$ and $N_{\rm HI}$. Color represents $T_{\rm s}$ of each point.
The dashed red lines and the dashed blue lines indicate $W_{\rm HI}$ derived with equations (4) and (6) for $T_{\rm s} = 10$\,K\,--\,100\,K and for $\tau_{\rm HI}$, 0.3, 1.0, 2.0 and 3.0, respectively.
Here $\Delta V$ in equations (4) and (6) is uniformly assumed to be 15\,km\,s$^{-1}$ . 
 \label{nhi_vs_whi}} 
\end{figure}

The observed $W_{\rm HI}$ as a function of the computed $N_{\rm HI}$ from equation (\ref{eq2+}) with colors showing the value of $T_{\rm s}=10$\,K\,--\,100\,K is shown in Figure \ref{nhi_vs_whi}. Generally, $W_{\rm HI}$ begins to saturate at H{\sc i} optical depth around 0.3. For lower $T_{\rm s}$, saturation begins at lower $N_{\rm HI}$, and for higher $T_{\rm s}$ the correlation between $W_{\rm HI}$ and $N_{\rm HI}$ becomes better than for lower $T_{\rm s}$. Since both $T_{\rm s}$ and $T_{\rm d}$ are determined by radiative heating and cooling (see Section 4 in Paper I), the qualitative trend of the $T_{\rm d}$-dependence of $W_{\rm HI}$ should be consistent with that found in Figure \ref{tau353_vs_whi}.

\begin{figure}
\epsscale{.8}
\plotone{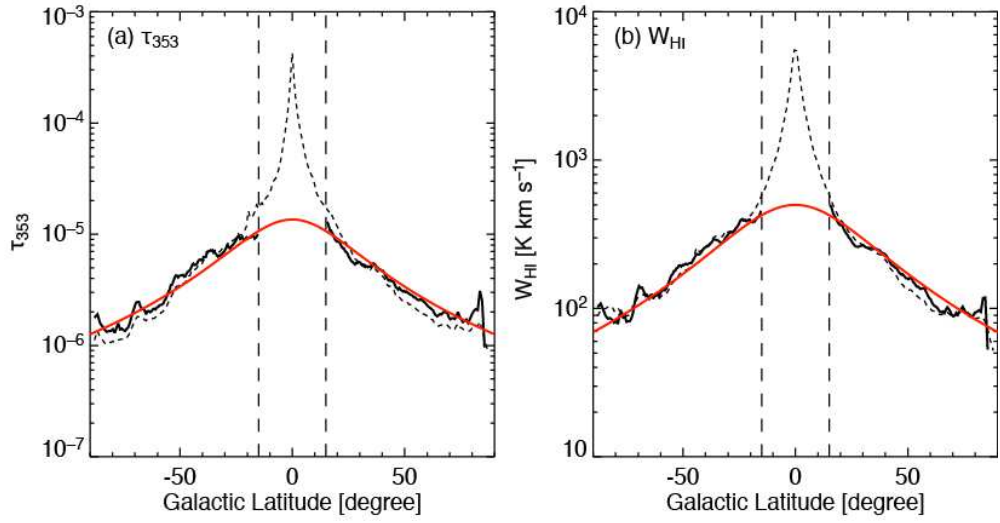}
\caption{Curves of averaged (a) $\tau_{353}$ and (b) $W_{\rm HI}$ along the Galactic latitude with mask (black solid line) and without mask (black dotted line). Vertical dashed lines indicate $|b|=15^\circ$. 
Red lines show the results of fitting with a Lorenzian function for the masked data. Half widths of the resulting curves are $28\fdg8$ and $36\fdg1$ for $\tau_{353}$ and $W_{\rm HI}$, respectively.
 \label{b_dist}} 
\end{figure}

We shall estimate the total ISM mass in the solar vicinity in the unmasked area. The H{\sc i} masses with optical-depth correction and that without correction are $1.0\times10^6$\,M$_\odot$ and $0.5\times10^6$\,M$_\odot$, respectively, for an assumed distance of 150\,pc (Paper I). This implies the mass increase due to the optical depth correction amounts to $0.5\times10^6$\,M$_\odot$, or a factor of $\sim$2.0, similar to the high-latitude clouds in Paper I. We also estimate the total ISM mass for the entire sky including the masked area by extrapolating $\tau_{353}$ and $W_{\rm HI}$. In Figure\,\ref{b_dist} we show the latitude distribution of $\tau_{353}$ and $W_{\rm HI}$ averaged in Galactic longitude, where the two curves, dashed and solid lines, indicate the entire area with and without masking, respectively. We fit the data by tentatively assuming a Lorenzian function as shown in Figure\,\ref{b_dist} to estimate the mass in the masked area. The estimated total masses with $N_{\rm HI}$ and $N_{HI}^*$ are $2.9\times10^6$\,M$_\odot$ and $1.2\times10^6$\,M$_\odot$, respectively, suggesting the optical depth correction amounts of $1.7\times10^6$\,M$_\odot$, or a factor of 2.5.

\section{Discussion: Ubiquitous optically-thick H{\sc i}, an alternative explanation for the dark gas}

The optically thick H{\sc i} gas has been identified in the region of MBM\,53, 54, 55 and HLCG\,92-35, and the CO clouds are enveloped by massive H{\sc i} gas having more than 10 times greater mass than the CO clouds (Paper I). The present study has shown that optically-thick H{\sc i} is common in interstellar space within 200\,pc of the Sun. The typical parameters of the H{\sc i} gas ($\geq70$\,\% of the total) are summarized as follows; $T_{\rm s}$\,=\,15\,K\,--\,35\,K, $\tau_{\rm HI}$\,=\,0.5\,--\,3.0, and $N_{\rm HI}=5\times10^{20}$\,cm$^{-2}$\,--\,$3\times10^{21}$\,cm$^{-2}$ (Figure \ref{hist_ts}). If we tentatively assume a typical line-of-sight depth of 5\,pc for the cold H{\sc i}, the average density is estimated to be 30\,cm$^{-3}$\,--\,190\,cm$^{-3}$. The ratio of the actual H{\sc i} column density $N_{\rm HI}$ to that obtained under the optically thin approximation $N_{\rm HI}$* is estimated to be 2.0 over the region analyzed. The high density and low temperature of the cold H{\sc i} gas are consistent with the temperature estimates in a model spherical cloud with the density range concerned, which is under heating by the interstellar radiation field and cooling by the atomic lines including C{\sc ii} \citep{2007ApJ...654..273G}.%

\begin{figure}
\epsscale{.7}
\plotone{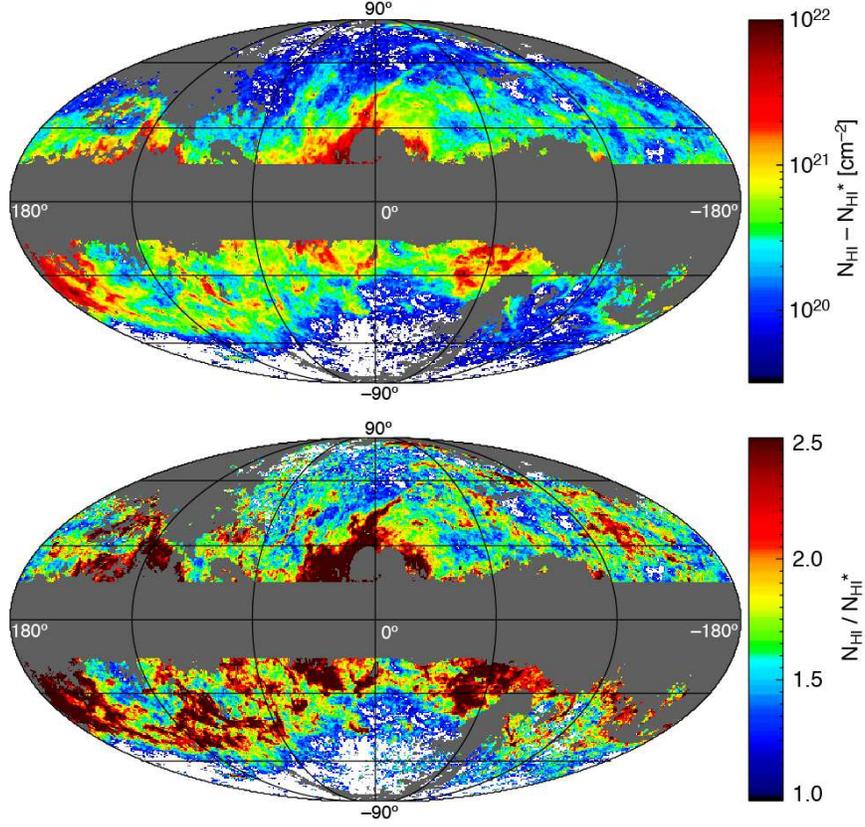}
\caption{
All-sky distributions of  (a) the subtracted column density of the cold H{\sc i}: $N_{\rm HI} - N_{\rm HI}*$, and of (b) the ratio of $N_{\rm HI}$ to $N_{\rm HI}*$. The masked region is shown in gray, and the points where $T_{\rm s}$ and $N_{\rm HI}$ are not determined are shown in white.
The center of the map is $(l, b)=(0^\circ, 0^\circ)$, and the coverages of l and b are from $-180^\circ$ to $+180^\circ$ and from $-90^\circ$ to $+90^\circ$, respectively. Dashed lines are plotted every $60^\circ$ in $l$ and every $30^\circ$ in $b$. 
 \label{darkgas_l0}} 
\end{figure}

\begin{figure}
\epsscale{.8}
\plotone{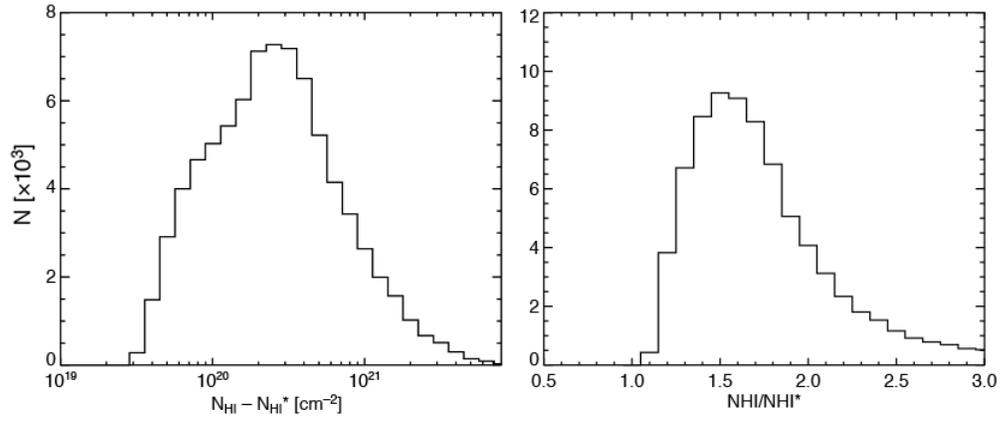}
\caption{
Histograms of (a) $N_{\rm HI}-N_{\rm HI}*$ and (b) $N_{\rm HI}/N_{\rm HI}*$ shown in Figure \ref{darkgas_l0}.
 \label{hist_darkgas}} 
\end{figure}

\citet{2005Sci...307.1292G} presented the dark gas based on $\gamma$-ray observations by EGRET, which is not ``detectable'' either by H{\sc i} or CO emission. Subsequently, \citet{2011A&A...536A..19P} discussed that the dust emission includes the dark gas component which has a similar distribution to the excess $\gamma$-rays. Dark gas is also seen in visual extinction \citep{2012A&A...543A.103P}. According to these studies, the distribution of the dark gas is largely similar to the CO gas, but it is spatially extended beyond the limit of CO detection. Its correlation with the H{\sc i} gas distribution did not seem to be strong in these previous studies that assumed optically thin H{\sc i}. 
The present analysis has shown that the H{\sc i} gas is dominated by an optically thick component. 
The typical H{\sc i} column density derived for the optically thick case, $\sim10^{21}$\,cm$^{-2}$, is consistent with that of the dark gas \citep{2005Sci...307.1292G}.
Figures \ref{darkgas_l0} and \ref{hist_darkgas} show $N_{\rm HI}-N_{\rm HI}$* and $N_{\rm HI}$/$N_{\rm HI}^*$ at 33$'$ resolution over the entire sky. 
These distributions are fairly similar to the dark gas distribution presented in \citet{2005Sci...307.1292G} and \citet{2011A&A...536A..19P}. More quantitative comparison between the optically thick H{\sc i} and $\gamma$-rays is a subject of a forthcoming paper.

\begin{figure}
\epsscale{0.9}
\plotone{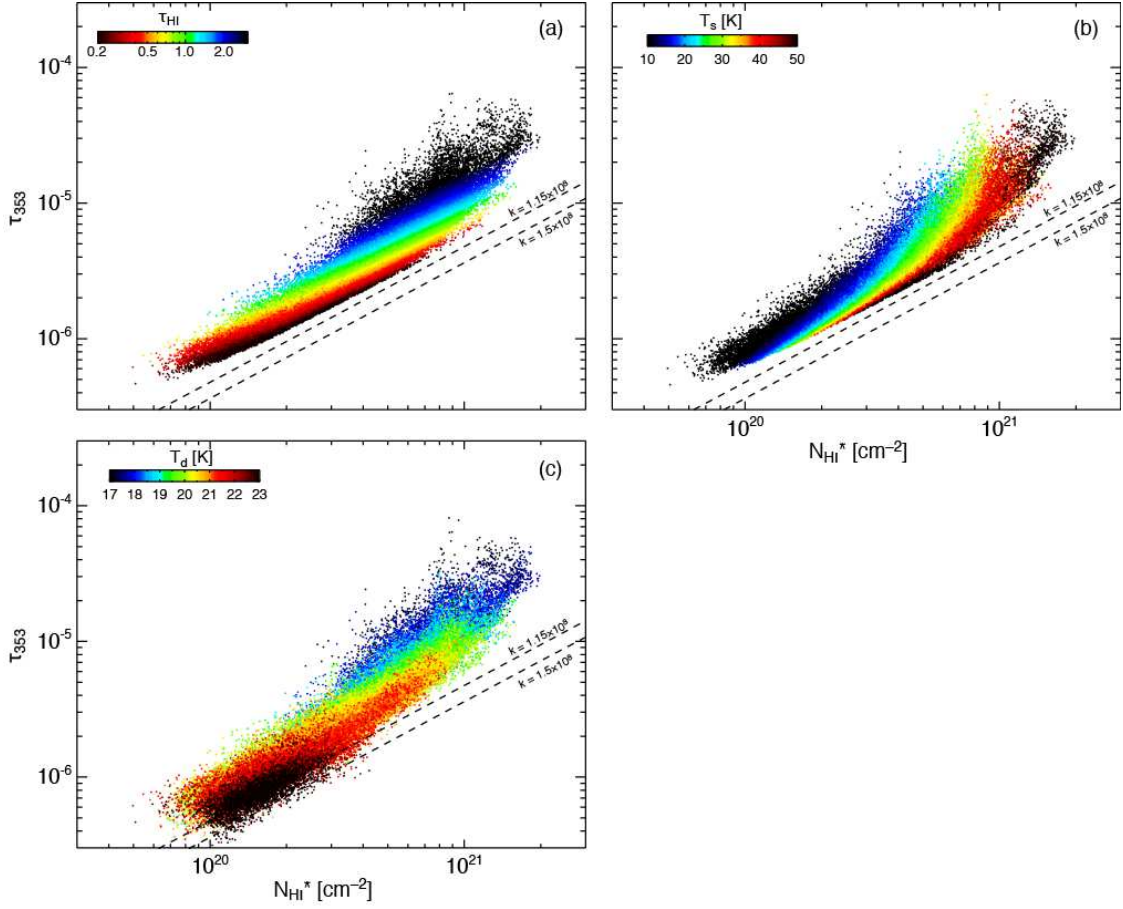}
\caption{Scatter plots between $N_{\rm HI}*$ and $\tau_{353}$. Color represents $\tau_{\rm HI}$, $T_{\rm s}$ and $T_{\rm d}$ in the panels (a), (b) and (c), respectively. The dashed lines indicate the relations for $k = 1.15\times10^8$\,K\,km\,s$^{-1}$ and $1.5\times10^8$\,K\,km\,s$^{-1}$.
 \label{nhi0_vs_tau353}} 
\end{figure}

\begin{figure}
\epsscale{0.7}
\plotone{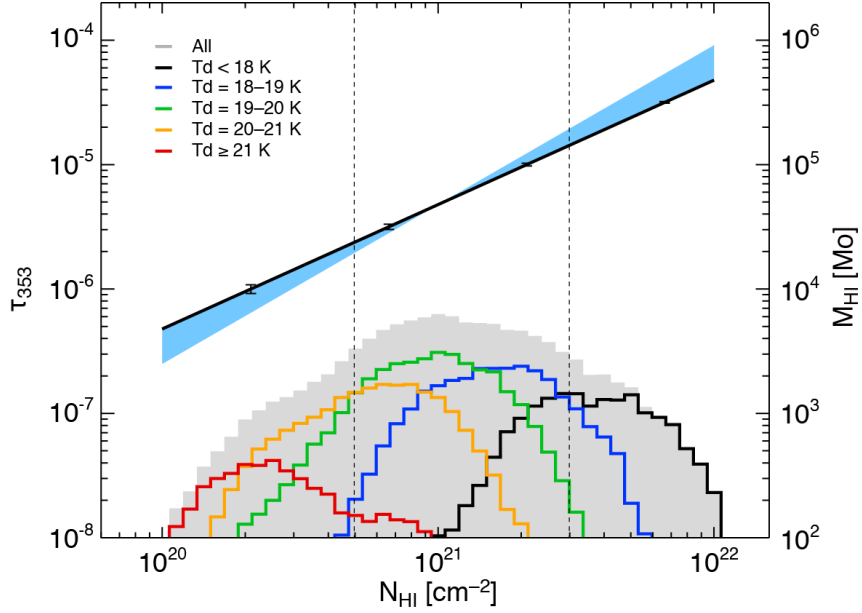}
\caption{
Correlation between $N_{\rm HI}$ and $\tau_{353}$ is shown by the thick line, where $N_{\rm HI}=2.1\times10^{26}\cdot \tau_{353}$. 
Representative values of the $\tau_{353}$ error are also plotted by bars.
The area colored in light-blue shows variation of the correlation provided by equation \ref{eq6} (See Section 4 for details).
Histograms show $N_{\rm HI}$ weighted by the H{\sc i} mass at various $T_{\rm d}$: black ($T_{\rm d} < 18$ K), blue (18 K $\leq T_{\rm d} < 19$ K), green (19 K $\leq T_{\rm d} < 20$ K), yellow (20 K $\leq T_{\rm d} < 21$ K) and red (21 K $\leq T_{\rm d}$). Gray includes all points. Here vertical dashed lines indicate the range of typical $N_{\rm HI}$ defined in Figure \ref{hist_ts}(c).
\label{nhi_vs_tau353}} 
\end{figure}

For more details, in Figure \ref{nhi0_vs_tau353} we show the scatter plot between $N_{\rm HI}^*$ and $\tau_{353}$, similar to Figure 6 in \citet{2011A&A...536A..19P}. Figures\,\ref{nhi0_vs_tau353}(a)--(c) show the relationship between $\tau_{353}$ and $N_{\rm HI}^*$ with colors indicating the dependence of the relationship on $\tau_{\rm HI}$, $T_{\rm s}$, and $T_{\rm d}$, respectively. These figures indicate that the apparent scatter in the plot reflects the difference in temperature ($T_{\rm d}$ or $T_{\rm s}$) or $\tau_{\rm HI}$, and that the actual scatter is much smaller than that in the $N_{\rm HI}^*$-$\tau_{353}$ correlation (Figure\,\ref{nhi0_vs_tau353}). According to the present analysis, $\tau_{\rm HI}$ takes maximum values of around 6\,--\,7 at $N_{\rm HI}^* \sim 10^{21}$ cm$^{-2}$ or $W_{\rm HI}\sim550$\,K\,km\,s$^{-1}$, and this optical depth effect causes the apparent bump at $N_{\rm HI}^*=4\times10^{20}$\,cm$^{-2}$\,--\,$2\times10^{21}$\,cm$^{-2}$ with a maximum at $N_{\rm HI}^* \cong 9\times10^{20}$\,cm$^{-2}$ as seen in Figure \ref{nhi0_vs_tau353}. The bump in the scatter plot is then interpreted in terms of the optical depth effect but not by the enhanced dispersion in the plot. We note that the bump is a natural outcome of the saturation effect in the optically thick H{\sc i} with no ad-hoc assumption, whereas it was ascribed to the property of the unknown dark gas in the previous interpretation \citep{2011A&A...536A..19P}. Figure \ref{nhi_vs_tau353} shows the correlation between the optical depth corrected H{\sc i} column density $N_{\rm HI}$ and $\tau_{353}$. Naturally, the scatter is small, on the order of 10 \% at maximum in Figure \ref{nhi0_vs_tau353}(c). We consider that Figure \ref{nhi_vs_tau353}  shows the real correlation with much smaller errors between $N_{\rm HI}$ and $\tau_{353}$. 

In Figure \ref{nhi0_vs_tau353}(c) we find that the points below the optically thin limit in a range of $N_{\rm HI}^*$ of $7\times10^{19}$\,cm$^{-2}$\,--\,$5\times10^{20}$\,cm$^{-2}$ are outside the regime where the present method has a solution. Most of these points are located at very high Galactic latitude higher than 60$^\circ$ as shown by black dots in Figures \ref{ts_allsky}(a) and (b), where $\tau_{353}$ in the sky is very low (like 10$^{-6}$), and are characterized by the highest $T_{\rm d}$ of larger than 23\,K; a typical point has $W_{\rm HI}\sim 80$\,K\,km\,s$^{-1}$ and $\tau_{353} \sim 7\times10^{-7}$. We suggest that the coefficient $k$ around $10^8$\,K\,km\,s$^{-1}$ in equation (\ref{eq2}) is larger in the higher $b$ (greater than 60$^\circ$) than in the lower $b$ (less than 60$^\circ$). If we adopt $k=1.5\times10^8$ K\,km\,s$^{-1}$ for instance, 99.7\,\% of all the data points are explained in the present scheme (Figure\,\ref{nhi0_vs_tau353}). This condition will be fulfilled if the dust optical depth is smaller by $\sim$30\,\% at such low column density. Such a trend is qualitatively consistent with smaller dust grains in the extremely low column density condition and is worth further study.

In order to evaluate the effect of the dust opacity which may depend on hydrogen column density, we shall test the following equation (\ref{eq6}) instead of equation (\ref{eq2+}); 

\begin{equation}
\left( \frac{N_{\rm HI}\,[\mathrm{cm^{-2}}]}{N_0} \right)^{1.28} = \frac{\tau_{353}}{\tau_0}. \label{eq6}
\end{equation}
This relation is derived by assuming the dependence of sub-mm dust optical depth which is proportional to $N_{\rm HI}^{0.28}$ for the total hydrogen column density $N_{\rm HI}$ above $10^{22}$ cm$^{-2}$ in Orion A \citep{2013ApJ...763...55R}. 
A trend of dust opacity evolution with column density is also recognized for clouds with column density of (3\,--\,7)$\times10^{21}$ cm$^{-2}$ in the Vela Molecular Ridge \citep{2012ApJ...751...28M} and in the Taurus filaments \citep{2013A&A...559A.133Y}. We made the same analysis of the cold H{\sc i} by using equation (\ref{eq6}) instead of equation (\ref{eq2+}) and have naturally found similar results to those above with some minor changes of physical quantities. Equation (\ref{eq6}) can alter $N_{\rm HI}$ by $-20$\,\%\,--\,$+30$\,\% at a typical range $5\times10^{20}$ cm$^{-2}$\,--\,$3\times10^{21}$\,cm$^{-2}$ as compared with the uniform dust opacity assumption, and the increase of the total H{\sc i} mass due to the H{\sc i} optical depth effect becomes slightly less by a factor of 2.1 instead of 2.0 for the unmasked area and 2.2 instead of 2.5 for the entire sky including the masked area (see Section 3.2). 
So it is not necessary to make a substantial change in the parameters of the cold H{\sc i}, even when the possibility of non-uniform dust opacity is taken into account.
Equation (\ref{eq6}) also suggests that the dust opacity variation is not the major cause of the observed poor correlation in Figure\,\ref{tau353_vs_whi}(a).
If the dust opacity variation is assumed to be substantial, the dust cross section must be increased to more than 300\,\% for $\tau_{\rm HI} = 3$, which is not consistent with equation (\ref{eq6}).

\begin{figure}
\epsscale{.7}
\plotone{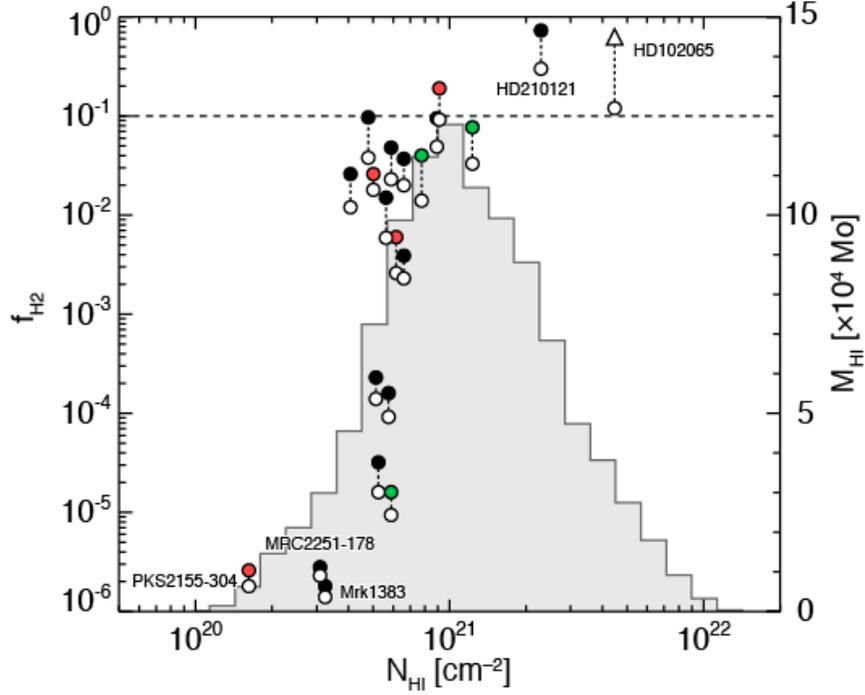}
\caption{
Plots of $f_{\rm H_2}$ for $N_{\rm HI}$. Filled circles and open circles show $f_{\rm H_2}$ before and after $\tau_{\rm HI}$ correction, respectively. 
The colors of the filled circles indicate different HPBWs of the 21-cm observations used in the UV studies. The filled black circles indicate a HPBW of 21$'$, and the green ones and the red ones have 9.7$'$, and larger than 30$'$, respectively. The open triangle indicates HD\,102065, whose H{\sc i} column density was directly measured with the $E(B-V)$ measurements without optically thin assumption \citep{2002ApJ...577..221R}.
Here the histogram of $N_{\rm HI}$ weighted by mass in Figure \ref{hist_ts}(c) is also shown. 
\label{figure_fuse}} 
\end{figure}

The present analysis has shown that cold H{\sc i} is a viable interpretation of the dark gas. It, however, does not exclude the possibility that H$_2$ is dominant instead of H{\sc i} in the dark gas. 
In order to explain the dark-gas origin, a possibility of molecular hydrogen with no CO emission has been discussed \citep[e.g.,][]{2011A&A...536A..19P}.
We here make a comparison with the results of direct UV absorption measurements of H$_2$ by FUSE and Copernicus \citep{2006ApJ...636..891G, 2002ApJ...577..221R, 2009ApJS..180..125R}. 
They observed about 80 lines of sight toward active galactic nuclei and galactic OB stars, and 21 of them are included in the present analysis. The 21 sources are taken from \citet{2006ApJ...636..891G} and \citet{2002ApJ...577..221R}, and are summarized in Table\,\ref{table_fuse}.
The H$_2$ abundance ratio is estimated by these authors as
\begin{equation}
 f_{\rm H_2}^* = \frac{2N_{\rm H_2}}{2N_{\rm H_2} + N_{\rm HI}^*},
\end{equation}
where $N_{\rm H_2}$ is the column density of H$_2$, and $N_{\rm HI}^*$ the H{\sc i} column density. It is to be noted that the $N_{\rm HI}^*$ in Table\,\ref{table_fuse} is derived under optically thin approximation measured with the 21-cm observations taken with multiple telescopes but not by the UV observations except for HD\,102065 whose H{\sc i} column density was measured with $E(B-V)$. The beam sizes (HPBWs) of the 21-cm observations were 9.7$'$\,--\,35$'$ as listed in Table\,\ref{table_fuse}. To make correction for the H{\sc i} optical depth, we replaced $N_{\rm HI}^*$ in equation (8) by the present $N_{\rm HI}$ as  
\begin{equation}
f_{\rm H_2} = \frac{2N_{\rm H_2}}{2N_{\rm H_2} + N_{\rm HI}}.
\end{equation}
$f_{\rm H_2}$ at each point is typically reduced by 30\,\%--\,80 \% from those in \citet{2006ApJ...636..891G} and \citet{2002ApJ...577..221R}, and the results are shown in Figure\,\ref{figure_fuse} and in Table \ref{table_fuse}. 
Figure \ref{figure_fuse} shows that molecular hydrogen is typically only $10^{-2}$\,--\,$10^{-1}$ or less of the total in the column density regime, $N_{\rm HI} \leq 1\times10^{21}$\,cm$^{-2}$, whereas there are only a few observations for $N_{\rm HI} > 10^{21}$\,cm$^{-2}$. 

\begin{figure}
\epsscale{.7}
\plotone{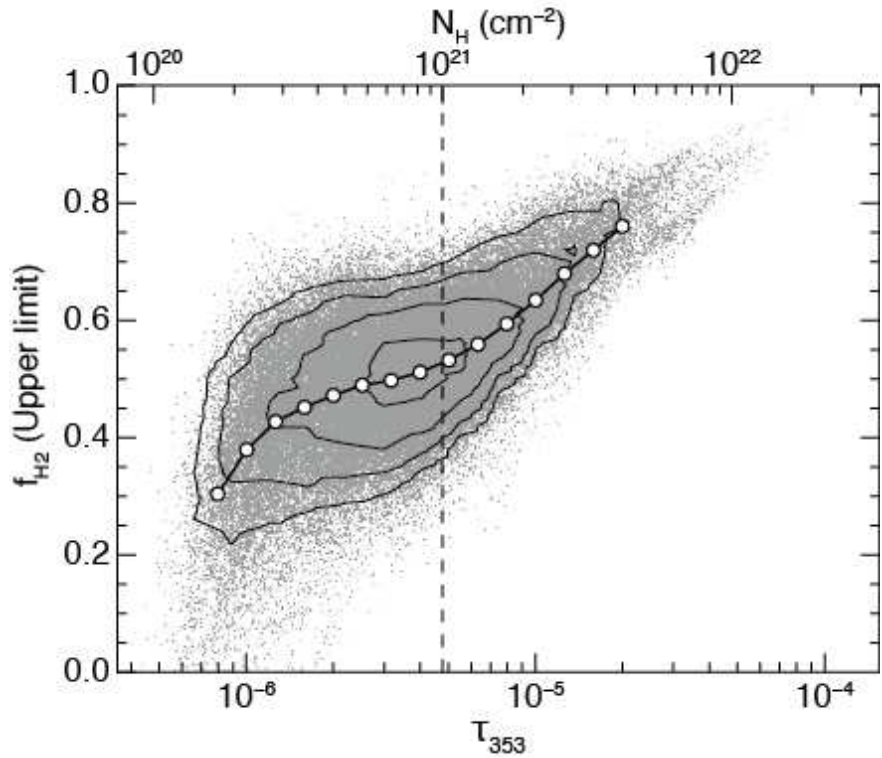}
\caption{
Scatter plots of $f_{\rm H_2}$(upper limit) as a function of $\tau_{353}$.  White circles indicate the average value of $f_{\rm H_2}$(upper limit) at each $\tau_{353}$. Contours are plotted at 80\,\%, 40\,\%, 20\,\%, and 10\,\% of the peak, and the vertical line is plotted at $N_{\rm H} = 1\times10^{21}$\,cm$^{-2}$ ($A_{\rm v}$ = 0.5 mag). 
\label{fh2}} 
\end{figure}

It is worthwhile to note that the minimum H{\sc i} column density is calculated from $W_{\rm HI}$ for the optically thin limit and is used to constrain the maximum $f_{\rm H_2}$ by assuming that equation (3) holds for H$_2$. 
For instance, at a data point having $W_{\rm HI}=280$\,K\,km\,s$^{-1}$ and $\tau_{353} =5.0\times10^{-6}$ in Figure 4 the minimum $N_{\rm HI}$ is calculated to be $5.1\times10^{20}$\,cm$^{-2}$ by equation (\ref{eq1}). 
This H{\sc i} column density corresponds to $\tau_{353}=2.4\times10^{-6}$ and the remaining $\tau_{353}=2.6\times10^{-6}$ gives the possible contribution of H$_2$ at maximum when only neutral hydrogen either atomic or molecular forms are considered. 
The upper limit for $f_{\rm H_2}$, $f_{\rm H_2}$(upper limit), is then calculated to be 0.52. In this way, we have calculated $f_{\rm H_2}$(upper limit) as a function of $\tau_{353}$ as shown in Figure\,\ref{fh2}. 
This provides a secure upper limit for $f_{\rm H_2}$, because a mixture of H{\sc i} and H$_2$ is more natural in the transition region between them, making the real $f_{\rm H_2}$ smaller than shown in Figure\,\ref{fh2}. 
Figure\,\ref{fh2} shows $f_{\rm H_2}{\rm (upper\ limit)}$ is typically $\sim$0.5, and the upper limit for the mass ratio of H$_2$ to H{\sc i} is 58\,\% in $N_{\rm H} = 10^{20}$\,cm$^{-2}$\,--\,$10^{22}$\,cm$^{-2}$.
For $N_{\rm H} < 1\times10^{21}$\,cm$^{-2}$ ($A_{\rm v}$ $<$ 0.5\,mag), $f_{\rm H_2}$(upper limit) is mostly less than 0.5 and the upper limit for the mass ratio of H$_2$ to H{\sc i} is estimated to be 48\,\% on average. 
As such, H$_2$ is nearly equal to or less than H{\sc i} in this regime with low extinction of $A_{\rm v}$ $<$ 0.5\,mag. 
This is consistent with Figure\,\ref{figure_fuse}, which indicates $f_{\rm H_2}$ is less than $10^{-2}$\,--\,$10^{-1}$. 
On the other hand, for $N_{\rm H} > 1\times10^{21}$\,cm$^{-2}$ ($A_{\rm v}$ $>$ 0.5\,mag) $f_{\rm H_2}$(upper limit) is 0.6 on average, suggesting that H$_2$ may dominate H{\sc i}, whereas H{\sc i} is still significant at a level of at least 10\,\% of the total hydrogen. 
The upper limit for the mass ratio of H{\sc i} to H$_2$ is estimated to be 64\,\% for $A_{\rm v}$ $>$ 0.5\,mag.

A recent study of the H{\sc i} at high latitude concluded that additional column density is provided by H$_2$ and that the H{\sc i} is not optically thick \citep{2014ApJ...783...17L}. This author found H{\sc i} intensity decreases as found in the present work, but the author rejected $T_{\rm s}$ around 30\,K for $E(B-V)\sim0.1$ mag to explain the intensity decrease. $E(B-V)\sim0.1$\,mag corresponds to $6\times10^{20}$\,cm$^{-2}$ in $N_{\rm HI}$. This reasoning is not justified because $T_{\rm s}$ can be as low as 30\,K even for $E(B-V)=0.1$ mag as shown by theoretical calculations of $T_{\rm s}$ as a function of visual extinction in a model cloud (see e.g., Figure 2 of Goldsmith et al. 2007; also for more extensive calculations see \citealt{1995ApJ...443..152W}).
The study by \citet{2014ApJ...783...17L} is therefore not justified as an counterargument for the optically thick H{\sc i}.
 It is also to be considered that the timescale of H$_2$ formation is considerably larger than the timescale of the local ISM as described below.

We shall here discuss theoretical aspects of the H{\sc i}-H$_2$ transition. The H$_2$ molecules are formed on the dust surface catalysis in the present-day universe and the H$_2$ formation timescale is given as $\sim10\cdot(n_{\rm HI}$ [cm$^{-3}]/100)$\,Myr \citep{1995ApJ...455..133H}. The current local clouds within 200\,pc of the Sun have a crossing timescale of $\leq$1\,Myr, which is too short to make H$_2$ as the major form by converting H{\sc i}, suggesting that H{\sc i} is generally overabundant relative to H$_2$. More detailed numerical simulations of H$_2$ formation from H{\sc i} gas have been undertaken by incorporating the H$_2$ formation reaction from purely H{\sc i} gas and it is shown for a density range of 10\,cm$^{-3}$\,--\,$10^3$\,cm$^{-3}$ that in $\sim$1\,Myr of cloud evolution the H$_2$ mass is about an order of magnitude smaller than the H{\sc i} mass and that even in $\sim$10\,Myr the H$_2$ mass is still dominated by the H{\sc i} mass \citep[e.g.,][]{2012ApJ...759...35I, 2012MNRAS.424.2599C}. 
It may be worthwhile to remark that Clark et al.'s paper presents ``CO-free H$_2$'', but their results actually show that H$_2$ is a minor component, corresponding to only $\sim$0.1 of H{\sc i} in mass (see their Figure 6). 
These results suggest that H$_2$ may not be the dominant form of the neutral hydrogen, while direct observations of H$_2$ to confirm this trend are not yet made at $N_{\rm H}$ larger than $10^{21}$\,cm$^{-2}$.

The dark gas in the Milky Way is observationally identified in the local space outside the Galactic plane where the line of sight contamination is not significant \citep{2005Sci...307.1292G, 2011A&A...536A..19P}. Therefore, the dark gas in the Milk Way is distributed within $\sim$200\,pc of the Sun in the Galactic latitude higher than 5$^\circ$. \citet{2010ApJ...716.1191W} presented numerical simulations of a giant molecular cloud with a lower density envelope and argued that an H$_2$ layer without CO surrounding a CO cloud may be significant in mass, providing a possible origin of the dark gas. The cloud size of a giant molecular cloud in \citet{2010ApJ...716.1191W} is much larger than that of the local clouds, and the timescale for a giant molecular cloud is as large as a few 10\,Myr \citep{1999PASJ...51..745F, 2009ApJS..184....1K, 2010ARA&A..48..547F}. These model simulations therefore do not apply to the local ISM having a much smaller timescale where the dark gas is identified.
Recent Herschel observations of C{\sc ii} toward the disk clouds at $|b|$ less than $1^\circ$ suggest that CO-free H$_2$ gas may be dominant in the disk outside the nearby regions analyzed in the present work \citep{2014A&A...561A.122L}. These observations also observed the giant molecular clouds that have ages of more than 10\,Myr and they do not apply to the local dark gas either. 

As a future direction an independent test of the H{\sc i} optical depth and $T_{\rm s}$ is possible by using the H{\sc i} absorption measurements toward extragalactic radio numerous continuum sources \citep{2003ApJS..145..329H, 2003ApJ...586.1067H, 2003ApJ...585..801D}.  We are able to extend this method to more continuum sources with a higher sensitivity and to compare the results with the present paper. 
Such measurements are also to be compared with numerical simulations of the H{\sc i}-H$_2$ transition, allowing us to have a deeper insight into the H$_2$ formation and the physical states of the hydrogen gas.

Another possibility which was not discussed in depth above is that the dust properties may be considerably different from the usual properties in the local space, as has been explored by the {\it Planck} collaboration \citep{2014A&A...571A..11P}. We shall defer to discuss this possibility until a full account of the {\it Planck} study is opened to the community.

\section{Conclusions}

We have carried out a study of the H{\sc i} gas properties in the local ISM by using dust properties derived from the {\it Planck}/{\it IRAS} all sky survey at sub-mm/far-infrared wavelengths. 
The H{\sc i} gas is in local regions within a few hundred pc of the Sun out of the Galactic plane, where giant molecular clouds do not exist.
We find $W_{\rm HI}$ shows poor correlation with the sub-mm dust optical depth $\tau_{353}$, whereas the correlation becomes significantly better if $T_{\rm d}$, ranging from 13\,K to 23\,K, is analyzed in several small ranges of width 0.5\,K. 
We hypothesize that the H{\sc i} is optically thick and the saturation of the H{\sc i} intensity is significant. We have shown that the H{\sc i} emission associated with the highest $T_{\rm d}$ shows a good correlation expressed by a linear regression and hence derive a relationship, $W_{\rm HI}$\ =\ $1.15\times10^8\cdot \tau_{353}$. An analysis of  $W_{\rm HI}$ and $N_{\rm HI}$ based on coupled equations of radiative transfer and the H{\sc i} optical depth yields both  $T_{\rm s}$ and $\tau_{\rm HI}$. $T_{\rm s}$ is typically in the range from 15\,K to 35\,K and $\tau_{\rm HI}$ from 0.5 to 3.0. The cold H{\sc i} gas typically has density of 30\,cm$^{-3}$\,--\,190\,cm$^{-3}$, $N_{\rm HI}$ = $5\times10^{20}$\,cm$^{-2}$\,--\,$3\times10^{21}$\,cm$^{-2}$, and $\Delta V_{\rm HI}\cong15$\,km\,s$^{-1}$. We argue that the ``dark gas'' is explained by cold H{\sc i} gas, which is 2\,--\,2.5 times more massive than the H{\sc i} gas derived under the optically thin approximation. We consider two alternative interpretations: one is that H$_2$ is dominant instead of H{\sc i}, and the other that variation of the dust opacity relative to the gas column density is significant. 
The fraction of H$_2$ $f_{\rm H_2}$ measured in the UV observations is consistent with that most of the hydrogen is atomic for $N_{\rm H}$ less than $1\times10^{21}$\,cm$^{-2}$, while for $N_{\rm H}$ larger than $1\times10^{21}$\,cm$^{-2}$ UV observations are only a few, insufficient to constrain $f_{\rm H_2}$. Minimum values of $N_{\rm HI}$ estimated by the optically thin limit constrain $f_{\rm H_2}$ to be less than $\sim$0.5, supporting that H{\sc i} is at least comparable to H$_2$. Theoretical studies of H{\sc i}-cloud evolution indicate that $f_{\rm H_2}$ is less than 0.1 for $\sim$1\,--\,Myr timescale by numerical simulations, lending support for that H{\sc i} dominates H$_2$ at density 10\,cm$^{-3}$\,--\,10$^3$\,cm$^{-3}$ in the local interstellar medium.
The second one the dust opacity variation is not reconciled with the general dust properties, either (equation 7).

$T_{\rm s}$ and $\tau_{\rm HI}$ cannot be disentangled by H{\sc i} intensity alone. 
This has been an obstacle in 21-cm H{\sc i} astronomy. The {\it Planck} dust optical depth offers a potential tool to disentangle this issue for an ISM column density range $10^{20}$\,cm$^{-2}$\,--\,$10^{22}$\,cm$^{-2}$. 
The opacity gives a measure of $N_{\rm HI}$ for given $T_{\rm d}$, if CO is not detectable and the background H{\sc i} gas is negligible. 
The present study suggests that the cold H{\sc i} is dominant in the local ISM and such cold H{\sc i} has important implications on related subjects, i.e., dust properties (in particular grain size evolution), the structure of molecular and atomic clouds, the interaction of H{\sc i} with cosmic rays, and the derivation of the $X_{\rm CO}$ factor. 
These issues will be subjects to be pursued in follow-up studies. 

\acknowledgments
We are grateful to John Dickey for his thoughtful comments and a valuable contribution on the H{\sc i} properties.
We are also grateful to Fraincois Boulanger and Jean-Phillippe Bernard for their initiative to begin the collaboration between the {\it Planck} and NANTEN2 teams.
This work was financially supported by Grants-in-Aid for Scientic Research (KAKENHI) of Japanese society for the Promotion of Science (JSPS) (grant numbers 24224005, 25287035, 23403001, 23540277, and 23740149-01). 
This work was also financially supported by the Young Research Overseas Visits Program for Vitalizing Brain Circulation (R2211) and the Institutional Program for Young Researcher Overseas Visits (R29) by the Japan Society for the Promotion of Science (JSPS) and by the grant-in-aid for Nagoya University Global COE Program, Quest for Fundamental Principles in the Universe: From Particles to the Solar System and the Cosmos," from MEXT. 
We acknowledge the use of the Legacy Archive for Microwave Background Data Analysis (LAMBDA), part of the High Energy Astrophysics Science Archive Center (HEASARC). HEASARC/LAMBDA is a service of the Astrophysics Science Division at the NASA Goddard Space Flight Center.
Some of the results in this paper have been derived using the HEALPix \citep{2005ApJ...622..759G} package.
We also utilize the data from Leiden/Argentine/Bonn Galactic H{\sc i} Survey and from WHAM, VTSS and SHASSA, combined by D. Finkbeiner (2003) to form an all-sky composite H$\alpha$ map, and the Virginia Tech Spectral-Line Survey (VTSS), which is supported by the National Science Foundation, and the Southern H-Alpha Sky Survey Atlas (SHASSA), which is supported by the National Science Foundation.
The Wisconsin H-Alpha Mapper is funded by the National Science Foundation.

\bibliographystyle{apj} 
%\bibliography{reference}

\begin{thebibliography}{}
\expandafter\ifx\csname natexlab\endcsname\relax\def\natexlab#1{#1}\fi

\bibitem[{{Braun}(2012)}]{2012ApJ...749...87B}
{Braun}, R. 2012, \apj, 749, 87

\bibitem[{{Calabretta} {et~al.}(2014){Calabretta}, {Staveley-Smith}, \&
  {Barnes}}]{2014PASA...31....7C}
{Calabretta}, M.~R., {Staveley-Smith}, L., \& {Barnes}, D.~G. 2014, \pasa, 31,
  7

\bibitem[{{Clark} {et~al.}(2012){Clark}, {Glover}, {Klessen}, \&
  {Bonnell}}]{2012MNRAS.424.2599C}
{Clark}, P.~C., {Glover}, S.~C.~O., {Klessen}, R.~S., \& {Bonnell}, I.~A. 2012,
  \mnras, 424, 2599

\bibitem[{{Dame} {et~al.}(2001){Dame}, {Hartmann}, \&
  {Thaddeus}}]{2001ApJ...547..792D}
{Dame}, T.~M., {Hartmann}, D., \& {Thaddeus}, P. 2001, \apj, 547, 792

\bibitem[{{Dickey}(2013)}]{2013pss5.book..549D}
{Dickey}, J.~M. 2013, {Galactic Neutral Hydrogen}, ed. T.~D. {Oswalt} \&
  G.~{Gilmore}, 549

\bibitem[{{Dickey} \& {Lockman}(1990)}]{1990ARA&A..28..215D}
{Dickey}, J.~M., \& {Lockman}, F.~J. 1990, \araa, 28, 215

\bibitem[{{Dickey} {et~al.}(2003){Dickey}, {McClure-Griffiths}, {Gaensler}, \&
  {Green}}]{2003ApJ...585..801D}
{Dickey}, J.~M., {McClure-Griffiths}, N.~M., {Gaensler}, B.~M., \& {Green},
  A.~J. 2003, \apj, 585, 801

\bibitem[{{Finkbeiner}(2003)}]{2003ApJS..146..407F}
{Finkbeiner}, D.~P. 2003, \apjs, 146, 407

\bibitem[{{Fukuda} {et~al.}(2014){Fukuda}, {Yoshiike}, {Sano}, {Torii},
  {Yamamoto}, {Acero}, \& {Fukui}}]{Fukuda2014}
{Fukuda}, T., {Yoshiike}, S., {Sano}, H., {et~al.} 2014, \apj, 788, 94

\bibitem[{{Fukui}(2013)}]{2013ASSP...34..249F}
{Fukui}, Y. 2013, in Astrophysics and Space Science Proceedings, Vol.~34,
  Cosmic Rays in Star-Forming Environments, ed. D.~F. {Torres} \& O.~{Reimer},
  249

\bibitem[{{Fukui} \& {Kawamura}(2010)}]{2010ARA&A..48..547F}
{Fukui}, Y., \& {Kawamura}, A. 2010, \araa, 48, 547

\bibitem[{{Fukui} {et~al.}(1999){Fukui}, {Mizuno}, {Yamaguchi}, {Mizuno},
  {Onishi}, {Ogawa}, {Yonekura}, {Kawamura}, {Tachihara}, {Xiao}, {Yamaguchi},
  {Hara}, {Hayakawa}, {Kato}, {Abe}, {Saito}, {Mano}, {Matsunaga}, {Mine},
  {Moriguchi}, {Aoyama}, {Asayama}, {Yoshikawa}, \&
  {Rubio}}]{1999PASJ...51..745F}
{Fukui}, Y., {Mizuno}, N., {Yamaguchi}, R., {et~al.} 1999, \pasj, 51, 745

\bibitem[{{Fukui} {et~al.}(2012){Fukui}, {Sano}, {Sato}, {Torii}, {Horachi},
  {Hayakawa}, {McClure-Griffiths}, {Rowell}, {Inoue}, {Inutsuka}, {Kawamura},
  {Yamamoto}, {Okuda}, {Mizuno}, {Onishi}, {Mizuno}, \&
  {Ogawa}}]{2012ApJ...746...82F}
{Fukui}, Y., {Sano}, H., {Sato}, J., {et~al.} 2012, \apj, 746, 82

\bibitem[{{Fukui} {et~al.}(2014){Fukui}, {Okamoto}, {Kaji}, {Yamamoto},
  {Torii}, {Hayakawa}, {Tachihara}, {Dickey}, {Okuda}, {Ohama}, {Kuroda}, \&
  {Kuwahara}}]{2014ApJ...796...59F}
{Fukui}, Y., {Okamoto}, R., {Kaji}, R., {et~al.} 2014, \apj, 796, 59, (Paper I)

\bibitem[{{Gillmon} {et~al.}(2006){Gillmon}, {Shull}, {Tumlinson}, \&
  {Danforth}}]{2006ApJ...636..891G}
{Gillmon}, K., {Shull}, J.~M., {Tumlinson}, J., \& {Danforth}, C. 2006, \apj,
  636, 891

\bibitem[{{Goldsmith} {et~al.}(2007){Goldsmith}, {Li}, \& {Kr{\v
  c}o}}]{2007ApJ...654..273G}
{Goldsmith}, P.~F., {Li}, D., \& {Kr{\v c}o}, M. 2007, \apj, 654, 273

\bibitem[{{G{\'o}rski} {et~al.}(2005){G{\'o}rski}, {Hivon}, {Banday},
  {Wandelt}, {Hansen}, {Reinecke}, \& {Bartelmann}}]{2005ApJ...622..759G}
{G{\'o}rski}, K.~M., {Hivon}, E., {Banday}, A.~J., {et~al.} 2005, \apj, 622,
  759

\bibitem[{{Grenier} {et~al.}(2005){Grenier}, {Casandjian}, \&
  {Terrier}}]{2005Sci...307.1292G}
{Grenier}, I.~A., {Casandjian}, J.-M., \& {Terrier}, R. 2005, Science, 307,
  1292

\bibitem[{{Heiles} \& {Troland}(2003{\natexlab{a}})}]{2003ApJS..145..329H}
{Heiles}, C., \& {Troland}, T.~H. 2003{\natexlab{a}}, \apjs, 145, 329

\bibitem[{{Heiles} \& {Troland}(2003{\natexlab{b}})}]{2003ApJ...586.1067H}
---. 2003{\natexlab{b}}, \apj, 586, 1067

\bibitem[{{Hollenbach} \& {Natta}(1995)}]{1995ApJ...455..133H}
{Hollenbach}, D., \& {Natta}, A. 1995, \apj, 455, 133

\bibitem[{{Inoue} \& {Inutsuka}(2012)}]{2012ApJ...759...35I}
{Inoue}, T., \& {Inutsuka}, S.-i. 2012, \apj, 759, 35

\bibitem[{{Kalberla} {et~al.}(2005){Kalberla}, {Burton}, {Hartmann}, {Arnal},
  {Bajaja}, {Morras}, \& {P{\"o}ppel}}]{2005A&A...440..775K}
{Kalberla}, P.~M.~W., {Burton}, W.~B., {Hartmann}, D., {et~al.} 2005, \aap,
  440, 775

\bibitem[{{Kalberla} \& {Kerp}(2009)}]{2009ARA&A..47...27K}
{Kalberla}, P.~M.~W., \& {Kerp}, J. 2009, \araa, 47, 27

\bibitem[{{Kawamura} {et~al.}(2009){Kawamura}, {Mizuno}, {Minamidani},
  {Filipovi{\'c}}, {Staveley-Smith}, {Kim}, {Mizuno}, {Onishi}, {Mizuno}, \&
  {Fukui}}]{2009ApJS..184....1K}
{Kawamura}, A., {Mizuno}, Y., {Minamidani}, T., {et~al.} 2009, \apjs, 184, 1

\bibitem[{{Langer} {et~al.}(2014){Langer}, {Velusamy}, {Pineda}, {Willacy}, \&
  {Goldsmith}}]{2014A&A...561A.122L}
{Langer}, W.~D., {Velusamy}, T., {Pineda}, J.~L., {Willacy}, K., \&
  {Goldsmith}, P.~F. 2014, \aap, 561, A122

\bibitem[{{Liszt}(2014)}]{2014ApJ...783...17L}
{Liszt}, H. 2014, \apj, 783, 17

\bibitem[{{Martin} {et~al.}(2012){Martin}, {Roy}, {Bontemps},
  {Miville-Desch{\^e}nes}, {Ade}, {Bock}, {Chapin}, {Devlin}, {Dicker},
  {Griffin}, {Gundersen}, {Halpern}, {Hargrave}, {Hughes}, {Klein}, {Marsden},
  {Mauskopf}, {Netterfield}, {Olmi}, {Patanchon}, {Rex}, {Scott}, {Semisch},
  {Truch}, {Tucker}, {Tucker}, {Viero}, \& {Wiebe}}]{2012ApJ...751...28M}
{Martin}, P.~G., {Roy}, A., {Bontemps}, S., {et~al.} 2012, \apj, 751, 28

\bibitem[{{Mohan} {et~al.}(2004){Mohan}, {Dwarakanath}, \&
  {Srinivasan}}]{2004JApA...25..185M}
{Mohan}, R., {Dwarakanath}, K.~S., \& {Srinivasan}, G. 2004, Journal of
  Astrophysics and Astronomy, 25, 185

\bibitem[{{Paradis} {et~al.}(2012){Paradis}, {Dobashi}, {Shimoikura},
  {Kawamura}, {Onishi}, {Fukui}, \& {Bernard}}]{2012A&A...543A.103P}
{Paradis}, D., {Dobashi}, K., {Shimoikura}, T., {et~al.} 2012, \aap, 543, A103

\bibitem[{{Planck Collaboration} {et~al.}(2011){Planck Collaboration},
  {Abergel}, {Ade}, {Aghanim}, {Arnaud}, {Ashdown}, {Aumont}, {Baccigalupi},
  {Balbi}, {Banday}, \& et~al.}]{2011A&A...536A..24P}
{Planck Collaboration}, {Abergel}, A., {Ade}, P.~A.~R., {et~al.} 2011, \aap,
  536, A24

\bibitem[{{Planck Collaboration} {et~al.}(2014{\natexlab{a}}){Planck
  Collaboration}, {Abergel}, {Ade}, {Aghanim}, {Alves}, {Aniano},
  {Armitage-Caplan}, {Arnaud}, {Ashdown}, {Atrio-Barandela}, \&
  et~al.}]{2014A&A...571A..11P}
---. 2014{\natexlab{a}}, \aap, 571, A11

\bibitem[{{Planck Collaboration} {et~al.}(2014{\natexlab{b}}){Planck
  Collaboration}, {Abergel}, {Ade}, {Aghanim}, {Alves}, {Aniano}, {Arnaud},
  {Ashdown}, {Aumont}, {Baccigalupi}, {Banday}, {Barreiro}, {Bartlett},
  {Battaner}, {Benabed}, {Benoit-L{\'e}vy}, {Bernard}, {Bersanelli},
  {Bielewicz}, {Bobin}, {Bonaldi}, {Bond}, {Bouchet}, {Boulanger}, {Burigana},
  {Cardoso}, {Catalano}, {Chamballu}, {Chiang}, {Christensen}, {Clements},
  {Colombi}, {Colombo}, {Couchot}, {Crill}, {Cuttaia}, {Danese}, {Davis}, {de
  Bernardis}, {de Rosa}, {de Zotti}, {Delabrouille}, {D{\'e}sert}, {Dickinson},
  {Diego}, {Dole}, {Donzelli}, {Dor{\'e}}, {Douspis}, {Dupac}, {Efstathiou},
  {En{\ss}lin}, {Eriksen}, {Falgarone}, {Finelli}, {Forni}, {Frailis},
  {Franceschi}, {Galeotta}, {Ganga}, {Ghosh}, {Giard}, {Giraud-H{\'e}raud},
  {Gonz{\'a}lez-Nuevo}, {G{\'o}rski}, {Gregorio}, {Gruppuso}, {Guillet},
  {Hansen}, {Harrison}, {Helou}, {Henrot-Versill{\'e}},
  {Hern{\'a}ndez-Monteagudo}, {Herranz}, {Hildebrandt}, {Hivon}, {Hobson},
  {Holmes}, {Hornstrup}, {Hovest}, {Huffenberger}, {Jaffe}, {Jaffe}, {Joncas},
  {Jones}, {Jones}, {Juvela}, {Kalberla}, {Keih{\"a}nen}, {Kerp}, {Keskitalo},
  {Kisner}, {Kneissl}, {Knoche}, {Kunz}, {Kurki-Suonio}, {Lagache},
  {L{\"a}hteenm{\"a}ki}, {Lamarre}, {Lasenby}, {Lawrence}, {Leonardi},
  {Levrier}, {Liguori}, {Lilje}, {Linden-V{\o}rnle}, {L{\'o}pez-Caniego},
  {Lubin}, {Mac{\'{\i}}as-P{\'e}rez}, {Maffei}, {Maino}, {Mandolesi}, {Maris},
  {Marshall}, {Martin}, {Mart{\'{\i}}nez-Gonz{\'a}lez}, {Masi}, {Massardi},
  {Matarrese}, {Mazzotta}, {Melchiorri}, {Mendes}, {Mennella}, {Migliaccio},
  {Mitra}, {Miville-Desch{\^e}nes}, {Moneti}, {Montier}, {Morgante},
  {Mortlock}, {Munshi}, {Murphy}, {Naselsky}, {Nati}, {Natoli}, {Noviello},
  {Novikov}, {Novikov}, {Oxborrow}, {Pagano}, {Pajot}, {Paoletti}, {Pasian},
  {Perdereau}, {Perotto}, {Perrotta}, {Piacentini}, {Piat}, {Pierpaoli},
  {Pietrobon}, {Plaszczynski}, {Pointecouteau}, {Polenta}, {Ponthieu}, {Popa},
  {Pratt}, {Prunet}, {Puget}, {Rachen}, {Reach}, {Rebolo}, {Reinecke},
  {Remazeilles}, {Renault}, {Ricciardi}, {Riller}, {Ristorcelli}, {Rocha},
  {Rosset}, {Roudier}, {Rusholme}, {Sandri}, {Savini}, {Spencer}, {Starck},
  {Sureau}, {Sutton}, {Suur-Uski}, {Sygnet}, {Tauber}, {Terenzi}, {Toffolatti},
  {Tomasi}, {Tristram}, {Tucci}, {Umana}, {Valenziano}, {Valiviita}, {Van
  Tent}, {Verstraete}, {Vielva}, {Villa}, {Wade}, {Wandelt}, {Winkel}, {Yvon},
  {Zacchei}, \& {Zonca}}]{2014A&A...566A..55P}
---. 2014{\natexlab{b}}, \aap, 566, A55

\bibitem[{{\textit{Planck} Collaboration} {et~al.}(2011){\textit{Planck}
  Collaboration}, {Ade}, {Aghanim}, {Arnaud}, {Ashdown}, {Aumont},
  {Baccigalupi}, {Balbi}, {Banday}, {Barreiro}, \&
  et~al.}]{2011A&A...536A..19P}
{\textit{Planck} Collaboration}, {Ade}, P.~A.~R., {Aghanim}, N., {et~al.} 2011,
  \aap, 536, A19

\bibitem[{{Rachford} {et~al.}(2002){Rachford}, {Snow}, {Tumlinson}, {Shull},
  {Blair}, {Ferlet}, {Friedman}, {Gry}, {Jenkins}, {Morton}, {Savage},
  {Sonnentrucker}, {Vidal-Madjar}, {Welty}, \& {York}}]{2002ApJ...577..221R}
{Rachford}, B.~L., {Snow}, T.~P., {Tumlinson}, J., {et~al.} 2002, \apj, 577,
  221

\bibitem[{{Rachford} {et~al.}(2009){Rachford}, {Snow}, {Destree}, {Ross},
  {Ferlet}, {Friedman}, {Gry}, {Jenkins}, {Morton}, {Savage}, {Shull},
  {Sonnentrucker}, {Tumlinson}, {Vidal-Madjar}, {Welty}, \&
  {York}}]{2009ApJS..180..125R}
{Rachford}, B.~L., {Snow}, T.~P., {Destree}, J.~D., {et~al.} 2009, \apjs, 180,
  125

\bibitem[{{Reich} \& {Reich}(1986)}]{1986A&AS...63..205R}
{Reich}, P., \& {Reich}, W. 1986, \aaps, 63, 205

\bibitem[{{Roy} {et~al.}(2013){Roy}, {Martin}, {Polychroni}, {Bontemps},
  {Abergel}, {Andr{\'e}}, {Arzoumanian}, {Di Francesco}, {Hill}, {Konyves},
  {Nguyen-Luong}, {Pezzuto}, {Schneider}, {Testi}, \&
  {White}}]{2013ApJ...763...55R}
{Roy}, A., {Martin}, P.~G., {Polychroni}, D., {et~al.} 2013, \apj, 763, 55

\bibitem[{{Strasser} \& {Taylor}(2004)}]{2004ApJ...603..560S}
{Strasser}, S., \& {Taylor}, A.~R. 2004, \apj, 603, 560

\bibitem[{{Torii} {et~al.}(2012){Torii}, {Fukui}, {Sano}, {Sato}, {Okuda},
  {Yamamoto}, {Kawamura}, {Mizuno}, {Onishi}, \& {Ogawa}}]{2012IAUS..284..389T}
{Torii}, K., {Fukui}, Y., {Sano}, H., {et~al.} 2012, in IAU Symposium, Vol.
  284, IAU Symposium, ed. R.~J. {Tuffs} \& C.~C. {Popescu}, 389--392

\bibitem[{{Wolfire} {et~al.}(2010){Wolfire}, {Hollenbach}, \&
  {McKee}}]{2010ApJ...716.1191W}
{Wolfire}, M.~G., {Hollenbach}, D., \& {McKee}, C.~F. 2010, \apj, 716, 1191

\bibitem[{{Wolfire} {et~al.}(1995){Wolfire}, {Hollenbach}, {McKee}, {Tielens},
  \& {Bakes}}]{1995ApJ...443..152W}
{Wolfire}, M.~G., {Hollenbach}, D., {McKee}, C.~F., {Tielens}, A.~G.~G.~M., \&
  {Bakes}, E.~L.~O. 1995, \apj, 443, 152

\bibitem[{{Ysard} {et~al.}(2013){Ysard}, {Abergel}, {Ristorcelli}, {Juvela},
  {Pagani}, {K{\"o}nyves}, {Spencer}, {White}, \&
  {Zavagno}}]{2013A&A...559A.133Y}
{Ysard}, N., {Abergel}, A., {Ristorcelli}, I., {et~al.} 2013, \aap, 559, A133

\end{thebibliography}

\clearpage

\begin{table}
\begin{center}
\caption{Fitting results for $k$ in Figure \ref{tau353_vs_whi}.}
\label{table_slope}
\begin{tabular}{ccccc}
\tableline\tableline
 $T_{\rm d}$ & \multirow{2}{*}{$N_{\rm pixel}$} &  Slope & Intercept &  \multirow{2}{*}{C.C.} \\
 ${\rm [K]}$ & & [K\,km\,s$^{-1}$] & [K\,km\,s$^{-1}$] & \\
 (1)&(2)&(3)&(4)&(5)\\
\tableline
$<$ 18.5 		& 9446	& $3.5\times10^7$	& 67.6	& 0.80 \\
18.5\,--\,19.0 	& 8327	& $5.1\times10^7$	& 27.9	& 0.85 \\
19.0\,--\,19.5 	& 12692	& $5.7\times10^7$	& 22.3	& 0.92 \\
19.5\,--\,20.0 	& 17233	& $6.4\times10^7$	& 13.8	& 0.93 \\
20.0\,--\,20.5 	& 17764	& $7.0\times10^7$	& 7.3		& 0.94 \\
20.5\,--\,21.0 	& 11299	& $7.7\times10^7$	& 4.1		& 0.94 \\
21.0\,--\,21.5 	& 5607	& $8.7\times10^7$	& $-1.3$	& 0.93 \\
21.5\,--\,22.0	& 3278	& $12.5\times10^7$	& $-14.3$	& 0.88 \\
22.0\,--\,22.5$^\dagger$	& 1693	& $10.2\times10^7$	& ---	& 0.77 \\
22.5 $\leq$$^\dagger$	& 3278	& $11.5\times10^7$	& ---	& 0.70 \\
\tableline
\end{tabular}
%% Any table notes must follow the \end{tabular} command.
\tablecomments{Column (1): $T_{\rm d}$ range. (2) Number of pixels used for the fitting. (3, 4) Fitting results. (5) Correlation coefficients for the plot. $^\dagger$ stands for the ranges in which intercept is fixed to be zero.}
\end{center}
\end{table}

\begin{table}
\begin{center}
\caption{Derived $\tau_{\rm HI}$ and $T_{\rm s}$ at the eight regions in Figure \ref{figure_fitting}.}
\label{table_fitting}
\begin{tabular}{ccccccccc}
\tableline\tableline
 \multirow{2}{*}{Region }	&	$l$ 	& $b$ &  $\tau_{353}$	& $T_{\rm d}$	& $W_{\rm HI}$	&	$N_{\rm HI}$	& $\tau_{\rm HI}$	&	$T_{\rm s}$  \\
 & [$^\circ$]	& [$^\circ$]	& [$\times10^{-6}$]	& [K]		& [K\,km\,s$^{-1}$] & [$\times10^{20}$\,cm$^{-2}$] &	& [K]	 \\
 (1)&(2)&(3)&(4)&(5)&(6)&(7)&(8)&(9)\\
\tableline

a	& 15.6	&  $45.6$		& 2.4		& 20.6	& 179.3	& 5.1		&$0.64^{+0.08}_{-0.08}$ 	& $21^{+2}_{-2}$\\
b	& 102.4	&  $-64.5$		& 1.8		& 20.5	& 180.8	& 3.9		&$0.25^{+0.11}_{-0.11}$ 	& $54^{+23}_{-12}$\\
c 	& 121.1	&  $-65.7$		& 2.3		& 20.3	& 197.5	& 4.7		&$0.40^{+0.08}_{-0.08}$ 	& $36^{+7}_{-5}$\\
d	& 162.6	&  $-56.3$		& 1.7		& 21.2	& 146.1	&3.7		& $0.43^{+0.07}_{-0.07}$ 	& $25^{+4}_{-3}$\\
e	& 173.2	&  $-29.3$		& 2.7		& 17.6	& 930.0	&5.7		& $3.06^{+0.13}_{-0.12}$ 	& $69^{+1}_{-1}$\\
f 	& 183.4	&  $17.7$		& 5.1		& 19.8	& 375.4	&6.3		& $0.23^{+0.06}_{-0.06}$ 	& $69^{+23}_{-13}$\\
g	& 252.7	&  $27.4$		& 3.3		& 21.0	& 250.1	&6.8		& $0.64^{+0.06}_{-0.06}$ 	& $28^{+2}_{-2}$\\
h 	& 325.5	&  $16.3$		& 6.8		& 20.4	& 386.7	&14.3	& $1.25^{+0.10}_{-0.11}$ 	& $20^{+1}_{-1}$\\

\tableline
\end{tabular}
%% Any table notes must follow the \end{tabular} command.
\tablecomments{Column (1): Name of region. (2, 3): Position in the galactic coordinate. (4--6): $\tau_{353}$, $T_{\rm d}$ and $W_{\rm HI}$ of the target region. (7): H{\sc i} column density without optically thin assumption. (8, 9): Derived $\tau_{\rm HI}$ and $T_{\rm s}$ with errors.}
\end{center}
\end{table}

\clearpage

\begin{table}
\begin{center}
\caption{Parameters of the $f_{\rm H_2}$ estimates.}
\scriptsize
\label{table_fuse}
\begin{tabular}{lccccccc}
\tableline\tableline
 \multirow{2}{*}{Target} & \multirow{2}{*}{($l$, $b$)} &\multirow{2}{*}{Ref.} & $N_{\rm H_2}$    & $N_{\rm HI}$ & HPBW & \multirow{2}{*}{$f_{\rm H_2}^*$}	& \multirow{2}{*}{$f_{\rm H_2}$} \\
&&&  [cm$^{-2}$]	 	& [cm$^{-2}$]	& [arcmin]		& 	& 	\\
(1)&(2)&(3)&(4)&(5)&(6)&(7)&(8)\\
\tableline
3C\,249.1		&($130\fdg39,\ 38\fdg55$)		&1	& $9.5\times10^{18}$	& $4.8\times10^{20}$	&21	& $9.8\times10^{-2}$		& $2.4\times10^{-2}$ \\
ESO\,141-G55	&($338\fdg18,\ -26\fdg71$)	&1	& $2.1\times10^{19}$	& $1.2\times10^{21}$	&34	& $7.8\times10^{-2}$		& $3.6\times10^{-2}$ \\
H1821+643	&($94\fdg00,\ 27\fdg42$)		&1	& $8.1\times10^{17}$	& $6.2\times10^{20}$	&9.7	& $6.0\times10^{-3}$		& $2.2\times10^{-3}$\\
HE\,1143-1810	&($281\fdg85,\ -41\fdg71$)	&1	& $3.5\times10^{16}$	& $5.1\times10^{20}$	&21	& $2.3\times10^{-4}$		& $1.4\times10^{-4}$ \\
MRC\,2251-178&($46\fdg20,\ -61\fdg33$)		&1	&$3.5\times10^{14}$		& $3.1\times10^{20}$	&21	& $2.8\times10^{-6}$		& $2.5\times10^{-6}$ \\
Mrk\,9		&($158\fdg36,\ 28\fdg75$)		&1	& $2.3\times10^{19}$	& $9.0\times10^{20}$	&21	& $9.5\times10^{-2}$		& $4.7\times10^{-2}$\\
Mrk\,335		&($108\fdg76,\ -41\fdg42$)	&1	& $6.8\times10^{18}$	& $5.9\times10^{20}$	&21	& $4.8\times10^{-2}$		& $1.9\times10^{-2}$\\
Mrk\,509		&($35\fdg97,\ -29\fdg86$)		&1	& $7.4\times10^{17}$	& $6.5\times10^{20}$	&21	& $3.9\times10^{-3}$		& $2.0\times10^{-3}$\\
Mrk\,1383		&($349\fdg22,\ 55\fdg12$)		&1	& $2.2\times10^{14}$	& $3.2\times10^{20}$	&21	& $1.8\times10^{-6}$		& $1.5\times10^{-6}$\\
Mrk\,1513		& ($63\fdg67,\ -29\fdg86$)		& 1	& $2.6\times10^{16}$	& $5.7\times10^{20}$	&21	& $1.6\times10^{-4}$		& $7.8\times10^{-5}$\\
MS\,0700.7+6338&($152\fdg47,\ 25\fdg63$)	&1	& $5.6\times10^{18}$	& $7.7\times10^{20}$	&35	& $4.0\times10^{-2}$		& $9.5\times10^{-3}$ \\
NGC\,7469	&($83\fdg10,\ -45\fdg47$)		&1	& $4.7\times10^{19}$	& $9.2\times10^{20}$	&9.7	& $1.9\times10^{-1}$		& $8.7\times10^{-2}$\\
PG\,0804+761	&($138\fdg28,\ 31\fdg03$)		&1	&$4.6\times10^{18}$		& $5.0\times10^{20}$	&9.7	& $2.6\times10^{-2}$		& $2.1\times10^{-2}$\\
PG\,0844+349	&($188\fdg56,\ 37\fdg97$)		&1	& $1.7\times10^{18}$	& $5.6\times10^{20}$	&21	& $1.5\times10^{-2}$		& $4.6\times10^{-3}$\\
PG\,1211+143	&($267\fdg55,\ 74\fdg32$)		&1	&$2.4\times10^{18}$		& $4.1\times10^{20}$	&21	& $2.6\times10^{-2}$		& $7.3\times10^{-3}$\\
PG\,1302-102	&($308\fdg59,\ 52\fdg25$)		&1	&$4.2\times10^{15}$		& $5.3\times10^{20}$	&21	& $3.2\times10^{-5}$		& $1.3\times10^{-5}$\\
PKS\,0558-504	&($257\fdg96,\ -28\fdg57$)	&1	&$2.8\times10^{15}$		& $5.9\times10^{20}$	&34	& $1.6\times10^{-5}$		& $1.0\times10^{-5}$ \\
PKS\,2155-304	&($17\fdg73,\ -52\fdg25$)		&1	&$1.5\times10^{14}$		& $1.6\times10^{20}$	&34	& $2.6\times10^{-6}$ 	& $1.5\times10^{-6}$\\
VII\,Zw\,118	&($151\fdg36,\ 25\fdg99$)		&1	&$6.9\times10^{18}$		& $6.6\times10^{20}$	&21	& $3.7\times10^{-2}$		& $1.9\times10^{-2}$\\
HD\,102065$^\dagger$	&($300\fdg03,\ -18\fdg00$)	&2	&$3.2\times10^{20}$		& $4.5\times10^{21}$	& ---	& $6.5\times10^{-1}$$^\dagger$ 	&	$1.2\times10^{-1}$ \\
HD\,210121	&($56\fdg88,\ -44\fdg46$)		&2	&$5.6\times10^{20}$		& $2.3\times10^{21}$	& 36	& $7.2\times10^{-1}$		& $2.5\times10^{-1}$ \\
\tableline
\end{tabular}
%% Any table notes must follow the \end{tabular} command.
\tablecomments{Column (1): Name of target. (3): Reference number. 1: \citet{2006ApJ...636..891G}, 2: \citet{2002ApJ...577..221R}. (4): $N_{\rm H_2}$ derived with the UV measurements. (5): $N_{\rm HI}$ obtained in the present study. (6) Beam size (HPBW) of the 21-cm observations used in the UV study. (7, 8) H$_2$ abundance ratios before and after $\tau_{\rm HI}$ correction. $^\dagger$$E(B-V)$ was used to directly measure the H{\sc i} column density without optically thin assumption in the literature.}
\end{center}
\end{table}

\end{document}